\documentclass[runningheads]{llncs}
\usepackage[T1]{fontenc}
\usepackage{graphicx}
\usepackage{listings}
\usepackage{xcolor}
\usepackage{amsmath}
\usepackage{amssymb}
\usepackage{tikz}
\usepackage{pgfplots}
\usepackage{array}
\usepackage{booktabs}
\usepackage{rotating}
\usepackage{longtable}
\usepackage{hyperref}
\usepackage{algorithm}
\usepackage{algpseudocode}
\usepackage{pifont}

\definecolor{codegray}{gray}{0.9}
\definecolor{rustpurple}{rgb}{0.58, 0.0, 0.83}
\definecolor{rustblue}{rgb}{0.25, 0.51, 0.85}
\definecolor{darkgreen}{rgb}{0.0, 0.75, 0.0}
\definecolor{amber}{rgb}{1.0, 0.75, 0.0}
\definecolor{darkred}{rgb}{0.75, 0.0, 0.0}

\lstdefinelanguage{Rust}{
  keywords={fn, pub, struct, let, mut, impl, for, match, if, else, while, loop, return, break, continue, crate, mod, enum, use, as, pub, super, self, Self, trait, const, static, ref, move, type, where, match, unsafe, extern, crate},
  keywordstyle=\color{rustpurple}\bfseries,
  ndkeywords={String, Vec, u32, i32, u64, i64, usize, isize, bool, Option, Result, Some, None, Ok, Err, Box, Rc, Arc},
  ndkeywordstyle=\color{rustblue}\bfseries,
  identifierstyle=\color{black},
  sensitive=true,
  comment=[l]{//},
  numbersep          = 3pt,
  morecomment=[s]{/*}{*/},
  commentstyle=\color{gray}\ttfamily,
  stringstyle=\color{brown}\ttfamily,
  morestring=[b]',
  morestring=[b]"
}

\lstdefinelanguage{quint}
{
, captionpos         = b
, numbers            = left
, numbersep          = 3pt
, stepnumber         = 1
, firstnumber        = auto
, numberbychapter    = true
, morecomment        = [l]{//}
, morecomment        = [l]{///}
, morecomment        = [s]{/*}{*/}
, morestring         = [b]"
, sensitive          = true
, escapeinside       = {@(}{)@}
, morekeywords       =
{ int, str, bool, set, seq,
Set, Map, List, Rec, Tup, not, and, or, iff, implies, exists, guess, forall, in, notin,
union, contains, fold, intersect, exclude, subseteq, map, applyTo, filter,
powerset, flatten, seqs, chooseSome, isFinite, cardinality, get, put, keys,
mapBy, setOfMaps, set, setBy, fieldNames, with, tuples, append, concat, head,
tail, length, nth, indices, replaceAt, slice, select, foldl, foldr, to, always,
eventually, next, orKeep, mustChange, enabled, weakFair, strongFair, guarantees,
existsConst, forallConst, chooseConst, if, else, match, all, any,
module, import, const, var, def, val, pure,
nondet, action, temporal, assume, type,
Bool, Int, Nat, false, true,
}
}

\lstdefinestyle{quintstyle}{
    backgroundcolor=\color{codegray},   
    commentstyle=\color{blue},
    keywordstyle=\color{purple},
    numberstyle=\tiny\color{gray},
    stringstyle=\color{brown},
    basicstyle=\ttfamily\footnotesize,
    breakatwhitespace=false,         
    breaklines=true,                 
    captionpos=b,                    
    keepspaces=true,                 
    numbers=left,                    
    numbersep=1pt,                  
    showspaces=false,                
    showstringspaces=false,
    showtabs=false,                  
    tabsize=2
}

\lstdefinestyle{ruststyle}{
    backgroundcolor=\color{codegray},   
    commentstyle=\color{gray},
    keywordstyle=\color{rustpurple},
    numberstyle=\tiny\color{gray},
    stringstyle=\color{brown},
    basicstyle=\ttfamily\footnotesize,
    breakatwhitespace=false,         
    breaklines=true,                 
    captionpos=b,                    
    keepspaces=true,                 
    numbers=left,                    
    numbersep=1pt,                  
    showspaces=false,                
    showstringspaces=false,
    showtabs=false,                  
    tabsize=2
}

\definecolor{pastelpurple}{rgb}{0.7, 0.6, 0.8} 
\definecolor{royalblue}{rgb}{0.25, 0.41, 0.88}    

\definecolor{templatekeyword}{rgb}{0.25, 0.41, 0.88}

\lstdefinestyle{templatestyle}{
  basicstyle=\ttfamily\footnotesize, 
  keywordstyle=\color{templatekeyword},
  backgroundcolor=\color{codegray},
  morekeywords={@@@NAME@@@,@@@DESCRIPTION@@@,@@@STUB@@@,@@@QUINT,TYPE,DEFINITIONS@@@,@@@CONSTANTS@@@,@@@DEC@@@,@@@QUINT,IMPORTS@@@,ERRORS@@@,@@@MESSAGE,HANDLER@@@,@@@IO,EXAMPLES@@@,@@@ORIGINAL,IMPLEMENTATION@@@},
  breaklines=true, 
  breakatwhitespace=true, 
  flexiblecolumns=true, 
  columns=fullflexible, 
  frame=single,
  breakindent=0pt,          
  captionpos=b,                    
}

\lstdefinestyle{descriptionstyle}{
    basicstyle=\ttfamily\footnotesize,
    keywordstyle=\color{templatekeyword},
    backgroundcolor=\color{codegray},
    breaklines=true,
    columns=fullflexible,
    frame=single,
    captionpos=b,                    
    xleftmargin=0pt,
    xrightmargin=0pt,
    framexleftmargin=0pt,
    framexrightmargin=0pt,
    breakindent=0pt,         
}

\lstdefinestyle{errorstyle}{
    basicstyle=\ttfamily\footnotesize,  
    frame=tb,                           
    framerule=0.5pt,                    
    framesep=5pt,                       
    showstringspaces=false,             
    breaklines=true,                    
    breakindent=0pt,          
    captionpos=b,                    
}

\usetikzlibrary{arrows, arrows.meta, automata, backgrounds, calc, decorations.pathreplacing, fadings, shadows.blur, shapes, fit, positioning, decorations.pathreplacing}
\usetikzlibrary{overlay-beamer-styles}
\tikzset{>=stealth, auto}
\tikzset{every state/.append style={minimum size=7mm}}
\tikzset{every edge/.append style={shorten >=1pt}}
\def\tikzshadowopacityregister{1}
\tikzset{
  invisible/.append code={\def\tikzshadowopacityregister{0}},
  every shadow/.style={opacity=\tikzshadowopacityregister}
}
\tikzset{every shadow/.append style={shadow xshift=1.5pt, shadow yshift=-1.5pt}}

\tikzset{neuron/.style={state}}
\tikzset{input-layer/.style={state, draw=darkgreen, fill=darkgreen!20}}
\tikzset{hidden-layer/.style={state, draw=structure.fg, fill=structure.fg!20}}
\tikzset{output-layer/.style={state, draw=darkred, fill=darkred!20}}

\pgfplotsset{compat=1.18}
\pgfplotsset{
    eq-query-style/.style={
		scale only axis=true,
		xmax=100,
		ymax=2500,
		axis x line=bottom,
		axis y line=left,
		title style={font=\scriptsize, at={(0.5,.8)}, above},
		label style={font=\footnotesize},
		xtick style={font=\scriptsize},
		ytick style={font=\scriptsize},
    }
}

\newcommand{\naturals}{\mathbb{N}}

\newcommand{\cmark}{\textcolor{darkgreen}{\ding{51}}} 
\newcommand{\mixed}{\textcolor{amber}{\textbf{?}}}    

\newcommand{\tlaplus}{TLA\textsuperscript{+}}

\newcommand{\todo}[1]{}

\newcommand{\mynote}[1]{}

\newcommand{\jan}[1]{#1}

\begin{document}
\title{Accessible Smart Contracts Verification:\\ Synthesizing Formal Models with Tamed LLMs}
\titlerunning{Synthesizing Formal Models with Tamed LLMs}
%
\author{Jan Corazza\inst{1}\orcidID{0009-0000-1342-0117} \and
Ivan Gavran\inst{2}\orcidID{0000-0002-2049-6264} \and
Gabriela Moreira\inst{2}\orcidID{0000-0003-0275-5717} \and
Daniel Neider\inst{1}\orcidID{0000-0001-9276-6342}}
\authorrunning{J. Corazza et al.}
%
\institute{Research Center Trustworthy Data Science and Security, TU Dortmund University, Dortmund, Germany\\
\email{\{jan.corazza,daniel.neider\}@tu-dortmund.de} \and
Informal Systems\\
\email{\{ivan,gabriela\}@informal.systems}}
\maketitle              
\begin{abstract}
When blockchain systems are said to be \emph{trustless}, what this really means is that all the trust is put into software.
Thus, there are strong incentives to ensure blockchain software is correct---vulnerabilities here cost millions and break businesses.
One of the most powerful ways of establishing software correctness is by using formal methods.
Approaches based on formal methods, however, induce a significant overhead in terms of time and expertise required to successfully employ them.
Our work addresses this critical disadvantage by automating the creation of a formal model---a mathematical abstraction of the software system---which is often a core task when employing formal methods.
We perform model synthesis in three phases: we first transpile the code into model stubs; then we ``fill in the blanks'' using a large language model (LLM); finally, we iteratively repair the generated model, on both syntactical and semantical level.
In this way, we significantly reduce the amount of time necessary to create formal models and increase accessibility of valuable software verification methods that rely on them.
The practical context of our work was reducing the time-to-value of using formal models for correctness audits of smart contracts.

\end{abstract}

\section{Introduction}

Modern blockchain platforms such as Ethereum and Cosmos extend the underlying consensus technology pioneered by Bitcoin towards a generalized computing platform.
Rather than solely reaching consensus on transactions modifying the state of a ledger, these platforms also enable consensus on program executions.
Roughly speaking, they are platforms for decentralized computing capable of executing stateful protocols---called smart contracts---in a trustless manner.
For example, one may write a smart contract for holding an auction without having to put one's trust into a centralized marketplace provider such as eBay.

Currently, the smart contract ecosystem manages assets worth billions of dollars.
This substantial value makes smart contracts prime targets for cyberattacks.
A successful exploit can result in catastrophic financial losses, potentially devastating businesses and the investments of numerous individuals.

Due to these high stakes, companies launching blockchain-based products have begun placing an increasingly high priority on carrying out thorough software audits before the launch.
Most auditors, however, base their work on manual line-by-line code inspection, foregoing more powerful tools such as formal methods.

Formal methods are a potent tool in the software verification toolkit, particularly for high-stakes environments like blockchain-based systems. 
They rely on rigorous logical reasoning to either automatically uncover vulnerabilities or assert their absence, thereby mathematically guaranteeing that certain types of bugs do not exist.
Not everything is perfect, however: for large codebases and expressive languages, formal methods scale poorly. 
One common way to address the problem of scalability is to build a formal model of the system and reason about that model instead of the full system implementation.
We call the techniques embracing this approach \emph{model-based techniques} (MBT).

A formal model is a mathematical abstraction of the software system---the art of modeling is about abstracting away all unnecessary details while preserving relevant aspects of the system.
For example, in a banking application, a model will capture how funds change owners.
When verifying the model, we can ensure that users cannot access funds that do not belong to them.

We can reason about properties of the model, and establish the connection of the model to the software using model-based testing~\cite{mbtSurvey} techniques.
Thus, model-based techniques provide an effective framework for verifying software. 
However, because of the initial cost of writing a model and the need for specialized knowledge of the modeling language, model-based techniques did not yet see widespread adoption.

In this work, we reduce the effort for an auditor writing a model, with help of large language models (LLMs).
We tame LLMs' propensity for hallucination by setting a hard frame around it through mechanical model synthesis, and through iterative repair of the generated model, until it conforms to the user's (partial) specification.

\subsubsection*{An auditor's workflow (model-based techniques)}
Let us examine the workflow of an auditor who attempts to use model-based techniques (we get into more details of those techniques in Section~\ref{subsec:mbt}).
After understanding the logic of the codebase, the auditor \textbf{models} it in a modeling language.
Now the auditor is able to reason about the model, but a connection to the implementation is missing.
Thus, the auditor \textbf{writes some glue code} (called an \emph{adapter}), which translates model's abstract traces into test cases.

These two steps are already time-consuming for an auditor already experienced in MBT.
For auditors just starting with MBT, they can also be daunting.
Given that audits are typically time-constrained, the time that needs to be invested is often used as an argument against using (otherwise effective and useful) model-based techniques.

Therefore, the primary objective of our work is to significantly reduce the time and effort needed to apply model-based techniques to smart contracts.
To this end, we propose an automated method that generates a formal model of a smart contract, based on the contract itself and the auditor's\footnote{We refer to an auditor as the main user of our method because our work was done in the context of a team performing correctness audits. The method is equally useful for developers aiming to employ formal methods in their work.} input.

\subsubsection*{Overview of the method}
Our method allows the auditor-user to automate parts of the MBT workflow, or take over in the middle of the process.
Importantly, at every step, the method provides a valuable artifact to the auditor.

As a first step, we transpile the smart contract(s) into a stub of the model: the stub captures interleavings of different actions, but does not capture the actions' semantics.
As a part of the transpilation process, we are also generating the stub of the adapter.
At this point, the auditor can use both stubs to run fuzzing tests against the contract.

In the second step, we ask the user to provide a natural language (NL) description of each action and a couple of input-output examples.
(An action is a transformation of the state, so the notion of I/O examples is well-defined.)
Our method then interacts with an LLM: each time the generated model is not aligned with the user-provided examples, it generates the prompt pointing out the inconsistency and requesting new generation.
Throughout our work we utilize OpenAI's Chat Completions API for this purpose, specifically GPT-4o.

The main insight on which we based the design of our tool is that the careful separation of concerns between mechanical synthesis (transpilation) and statistical synthesis (LLM-based) is crucial.
Relying on an LLM to generate the whole model results in hallucinations (e.g., using non-existent helper functions) or convincingly looking yet wrong models.
Conversely, transpiling the full contract into a model is hopeless: even if it worked, it would result in line-by-line re-implementation of the contract, instead of a good abstraction.
We suggest \emph{taming} as an appropriate metaphor: model stubs give a firm frame that constraints the wild creative power of LLMs.

\begin{figure}[t]
    \centering
    \input{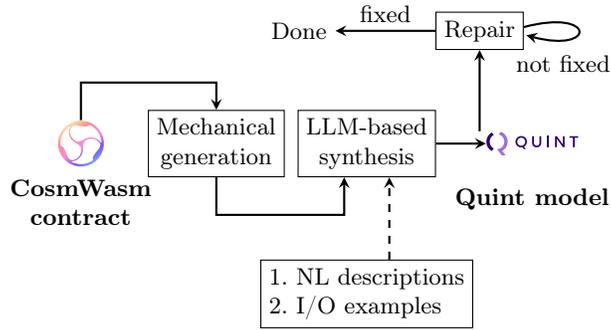}
    \caption{High-level illustration of the method for generating a Quint model from a CosmWasm contract.}
    \label{fig:method}
\end{figure}

We focus on smart contracts implemented using CosmWasm~\cite{cosmWasm}, a smart-contract platform of the Cosmos ecosystem.
CosmWasm contracts are written in Rust, and the models that our tool generates are written in Quint~\cite{quint}, an executable modeling language based on TLA---temporal logic of actions~\cite{tla}.
Figure~\ref{fig:method} provides a high-level overview of our method for transforming a CosmWasm contract into a Quint model.
The process involves three main steps.

\begin{enumerate}
\item \textbf{Mechanical Generation}: Transpile as much of the Rust source code into Quint, leaving placeholders (``stubs'') for parts that cannot be easily transpiled.
\item \textbf{LLM-based Synthesis}: Use a large language model (LLM) to complete the stubs based on user's input, thereby finishing the Quint model.
\item \textbf{Testing and Repair}: Conduct a series of tests on the generated model and identify the LLM's syntactic and semantic hallucinations. Then, provide the information about the problem to the LLM and ask for it to be fixed.
\end{enumerate}


In Sections \ref{sec:related-work} and \ref{sec:preliminaries}, we discuss related work and provide the necessary background information, respectively.
In Section~\ref{sec:model-generation}, we elaborate the technical details involved in each step of our method.

To evaluate its effectiveness, we created a benchmark based on CosmWasm CTF~\cite{ctf}, a collection of publicly available CosmWasm smart contracts with injected bugs.
Along with the evaluation results on the CTF benchmark, we also conducted two case studies of using our method as a part of an auditor's work.
The evaluation results (described in Section~\ref{sec:evaluation}) suggest the following conclusion: our method successfully generates models for the benchmark contracts and is useful for auditors.
Using the method, we uncovered a bug in a real-world smart contract.




\section{Related Work}
\label{sec:related-work}

In this section, we provide an overview of existing research and technologies related to formal methods for blockchain systems and LLM code generation, and place our work within this broader context.

\subsection{Formal Methods for Blockchain Systems}

Software vulnerabilities in the blockchain world imply high financial and reputational loss.
Once on-chain, the software with a fix cannot be deployed quickly.
Therefore, there has been a lot of effort in ensuring blockchain systems are correct.

On the level of the underlying blockchain infrastructure, engaging software verification experts to assess the correctness of the critical parts has become a standard. 
The experts employ a variety of tools to conduct their analysis, for instance: Jepsen~\cite{jepsen} for fault tolerance, modeling with \tlaplus{}~\cite{tlaPlus} for proving properties of inter-blockchain communication~\cite{ibcSpecs}, or providing a reference implementation~\cite{ethReference} in Dafny~\cite{dafny}.

Verifying individual smart contracts is (in principle, at least) a more attainable task than verifying the infrastructure, while the incentives remain equally high.
Thus, there is abundance of tools for smart contracts correctness assurance. Examples include Certora~\cite{certora},  Slither~\cite{slither}, Mythril~\cite{mythril}, Echidna~\cite{echidnaPaper}, Quint~\cite{quint}, to name but a few.

One thing these tools all share is that only specialized experts can use them.
The problem is well recognized and there is a trend for tools to get closer to users by integrating with commonly used frameworks. This approach is taken by e.g., Kontrol~\cite{kontrol} and Halmos~\cite{halmos}.

In this work, we take a different approach: we bring a modeling language closer to the users with help of automated synthesis. 
Such an approach could be applied to other formal verification tools as well.

\subsection{LLM Code Generation and Verification}

LLMs optimized for code generation showed stunning results~\cite{copilot,codeLlama}.
One concern when using LLMs for generating code is that they are inherently stochastic and offer no assurance that the generated code will be error-free.

Techniques such as PICARD~\cite{DBLP:journals/corr/abs-2109-05093} and Synchromesh~\cite{poesia2022synchromesh} utilize parsers to sequentially filter out undesirable tokens at inference time.
Synchromesh additionally introduces Target Similarity Tuning (TST), a mechanism for selecting few-shot demonstrations for in-context learning.
While TST is highly relevant for all techniques which rely on few-shot prompting, we did not incorporate it in our approach due to the small number of few-shot demonstrations at our disposal and generally satisfactory performance without it.
We discuss our prompting methodology and approach to LLM-based code synthesis in Section~\ref{sec:few-shot}.

At their core, these methods enforce syntactic and a limited number of semantic rules by restricting the token set the LLM can sample from, based on a provided syntax.
However, the rules enforceable at the parser level are limited and require an expert to encode them.
Furthermore, while correct-by-construction synthesis is a valuable ideal, we found that code generation capabilities of advanced LLMs have progressed to a point where existing inference-time approaches, especially those that rely on syntax level guarantees, are becoming less relevant.

Techniques like Clover~\cite{sun2024clover} cross-verify multiple LLM-generated artifacts, including code, documentation, and formal specifications. 
This bears similarity to our approach, in which the LLM generator gets feedback from other generated parts (discussed in more detail in Section~\ref{sec:repair}).
A radically different approach is taken by Sun et al.~\cite{sun2024ai}, who experiment with changing a programming language's grammar to make it a better fit for LLM-based generation.

There has been a lot of interesting work with promising results combining aspects of formal methods and LLM code generation~\cite{meng2024large,charalambous2023new,ye2020optimal,DBLP:conf/icse/JainVINPR022,misu2024towards,wang2024theoremllama}.
Our work fits into that broad context.

\section{Preliminaries}
\label{sec:preliminaries}

In this section, we cover the necessary background on model-based techniques, the Quint modeling language, and CosmWasm smart contracts, which form the primary focus of our work.

\subsection{Model-based Techniques}
\label{subsec:mbt}
In the context of this paper, a \emph{model} is an abstract, formal representation of a software system that can be automatically analyzed. 
Common examples include 1) writing models in the \tlaplus{}~\cite{tlaPlus} modeling language and analyzing them either by an explicit model checker TLC~\cite{tlc} or by a symbolic model checker Apalache~\cite{apalache}; and 2) writing models in the Alloy modeling language and analyzing them with the Alloy analyzer~\cite{alloy}.
In this paper, we focus on models written in Quint (see Section~\ref{subsec:quint} for a brief introduction to Quint).

Models are in general the most useful when modeling a (distributed) protocol \emph{before} the implementation.
Then, verifying the model can point to protocol problems early in the process (or assure us that the approach makes sense).
In our auditing practice, we most often come in once the software is already implemented, and we focus on finding bugs.
We can use models to help us find bugs in two ways: by verifying the \emph{code-based} model, or by generating tests from the \emph{ideal-based} model.

In the code-based approach, we write our model to be as close as possible to the implementation.
Then, finding violations of invariants with respect to the model will correspond to finding bugs in the code.
Of course, the pending question is how we can be sure that our model corresponds to the code: this is ensured by running model-based tests. 
The model generates traces, which are then executed against the implementation to verify conformance. 

A symmetrical situation occurs in the ideal-based approach.
There, we write a model to correspond to what we understand to be the correct functioning of the software.
From the model, we generate traces which are executed against the implementation and after each step of the trace, the state of the model is compared to the state of the implementation.
If there is a mismatch, it suggests there is a bug in the implementation.
A similar validity question is pending here, too: how can we be sure that our model is correct?
We can be assured of the model's correctness by writing and verifying the invariants that are expected to hold.

These two approaches represent somewhat extreme cases.
In reality, a model, tests, and invariants may co-evolve as our understanding of the system improves.
Furthermore, as noted already by many practitioners, one less tangible value of writing models is that it makes our thinking about the (audited) software much more precise. 
Thus, it is very common that bugs are recognized while writing the model, before any automatic analysis is run.

\subsection{Quint}
\label{subsec:quint}
Quint~\cite{quint} is a modeling language based on \tlaplus{}~\cite{tlaPlus}.
Quint's main advantage compared to \tlaplus{} is being significantly more in line with standard development practice.

It was designed according to the following principles:
\begin{itemize}
    \item Whenever a language concept has a ``standard'' syntax in mainstream languages, Quint adopts it. Conversely, when a concept is specific to modeling languages, Quint's syntax makes it explicit. As examples: Quint expresses inequality with \texttt{!=} (as opposed to \texttt{\#} and \texttt{/=}, used in \tlaplus{}); when a value needs to be chosen nondeterministically, Quint demands a \texttt{nondet} modifier.
    \item Users can specify types, and types are checked as soon as possible, identifying mistakes sooner rather than later.
    \item Indentation is encouraged with IDE tools, but does not carry meaning.
    \item There is a clear separation between functions that modify the model state, and those that do not.
    \item One can interact with parts of the model (i.e., in a REPL), before the whole model is finished.
    \item Models written in Quint remain compatible with \tlaplus{} and transpilation is possible, so all existing tools for \tlaplus{} can be used for Quint.
\end{itemize}

While most of these points may seem straightforward, the accepted wisdom among \tlaplus{} users was that a modeling language does not need to follow modern programming practices. 
Even more, that trying to stay sufficiently away from programming languages' mores is beneficial for modeling~\cite{tlaPlus}.

In our experience, Quint's principles add value by integrating modeling into the development process.
Quint's intuitive language and support for integration into standard, ergonomic tools have been essential for bringing modeling into practice.

\subsection{CosmWasm Smart Contracts}

CosmWasm~\cite{cosmWasm} (CW) is a smart contract platform of the Cosmos ecosystem. 
The name comes from the fact that CW contracts are compiled into WebAssembly (Wasm), enabling CW contracts to be written in multiple languages (the de facto standard programming language for CW at this point is Rust).

A CosmWasm smart contract consists of type definitions which define how the contract's data is stored on the blockchain, and entry-point implementations that operate on that data based on the messages that the contract receives.
Multiple contracts can communicate by sending each other messages, which are all delivered and processed as part of the same transaction as the initial user's message.

\section{Model Generation}
\label{sec:model-generation}

As stated in the introduction, our goal is to generate a Quint model for a given smart contract in three steps.
In the first step, which we call mechanical generation, we use the contract's Rust source code to determine the overall structure of Quint model.
This enables us to generate data types, function signatures, and parts of code that follow common CosmWasm patterns, such as receiving and replying to messages and mutating the contract's internal state.
However, we leave placeholders, which we refer to as stubs, in place of functions that encode the contract's main logic.
Filling these placeholders mechanically would require one to develop a complete Rust-to-Quint transpiler, including translation of side effects and loops, which is an exceedingly difficult task.
Additionally, mechanical generation would necessarily follow the Rust code by the letter, failing to sufficiently abstract away from it.
(In Figure~\ref{fig:listings-stubs} in Appendix~\ref{sec:additional-results}, we provide an example of a stub along with the original Rust code that induced it.)

Instead, in the second step, we prompt the LLM to complete the missing parts of the model.
In order to help the LLM accurately complete this task, we provide it with general information about Quint, user-provided NL descriptions, and I/O specifications.
We also include relevant parts of the contract's Rust implementation along with results of step 1 (primarily, stubs' signatures that the synthesized code must adhere to).
Most importantly, we provide the LLM with few-shot examples which demonstrate how to generate Quint code.

Finally, in the third step, we conduct a series of tests on the generated model to ensure its correctness, and again employ the LLM to repair any identified errors based on the error details.

\subsection{Mechanical Generation for Stubs}
\label{sec:mechanical-generation}

The mechanical generation phase analyzes a CosmWasm contract written in Rust and produces a corresponding Quint model with the overall structure, without attempting to generally translate Rust expressions. It introduces placeholders for the core elements that it cannot translate. It also generates an adapter test file in Rust, which reads traces produced by the generated model and replays them against the implementation. 
The tool is called \texttt{cosmwasm-to-quint} and the source code is available on GitHub\footnote{\url{https://github.com/informalsystems/cosmwasm-to-quint}}.

This section touches on the technical aspects of Rust tooling and their application in the generation process.
Readers unfamiliar with these details may choose to skip it, or refer to the Rust documentation\footnote{\url{https://doc.rust-lang.org/}}.
Furthermore, it also touches on the structure of Quint models.
Appendix~\ref{sec:quint-model-structure} provides a quick primer on the basic structure of Quint models.

There are several ways to work with Rust code, from building a new parser to using existing translation libraries.
After considering a few of those options, we chose a tool called \texttt{rustc\_plugin}~\cite{rustcPlugin} for that purpose.
\texttt{rustc\_plugin} is a framework that makes it easier to add callbacks to compilation phases of rustc (the Rust compiler), where all the information produced by the compiler is available. 

Since the compiler is capable of understanding the project's configuration (\texttt{Cargo.toml} and related files), the tool can be run for the entire project (as opposed to individual Rust files), which is a great convenience as CosmWasm projects often involve multiple files.
Some projects define more than one contract, each one in its own Rust crate.
The generation tool will generate a different Quint file and Rust test for each crate.
A significant constraint that affects the transpilation at this time is the lack of import resolution for external crates.
A potential solution for this is to download the external crates and compile them as well, so the definitions are compiled and compilation callbacks are called for them.
This is not currently automated by the tool, which means any definitions from external crates will be undefined in the Quint model (and therefore, name resolution for Quint will report errors).


\subsubsection*{Finding the contract state}
CosmWasm contracts have a state, and it has to be translated into a Quint state variable in order to be persisted between different transitions.
Listings~\ref{lst:rust}~and~\ref{lst:quint} show the input and the output of this process.
The most complex part is finding which definitions are related to the state definition, and expressing their types (as Quint state variables require a precise type annotation).
To this end, the tool scans all Rust constants and selects them as state variables if they satisfy all of the following conditions:
\begin{enumerate}
    \item the body of the constant is a call to a type with a relative path that resolves into a definition;
    \item the resolved definition is from a crate called ``cw\-\_storage\-\_plus''; and
    \item the resolved definition is either ``Map::new'' or ``Item::new''.
\end{enumerate}

If ``Map::new'', the tool looks for two generic arguments to determine the type of the keys and the values of the map, and produces a Quint type from them.
If ``Item::new'', the tool looks for a single generic argument, which is translated into the corresponding Quint type.

After collecting all state-related variable names and types, the generator will produce a Quint type called \texttt{ContractState}, defined as a record where field names are the variable names, and field types are their corresponding types.
This is more convenient than generating one state variable for each of the found variables because it is necessary to pass this to many functions, so having the information stored in a single value makes it easier.
The tool will always produce a state variable definition called \texttt{contract\_state} with this type, and also a value definition called \texttt{init\_contract\_state} with the default values with which to initialize the contract state.

\begin{lstlisting}[language=Rust, style=ruststyle, caption={Original CosmWasm state definition}, label={lst:rust}]
pub const LAST_ID: Item<u64> = Item::new("lock_id");
pub const LOCKUPS: Map<u64, Lockup> = Map::new("lockups");
\end{lstlisting}

\begin{lstlisting}[language=quint, style=quintstyle, caption={Generated Quint code for the contract state}, label={lst:quint}]
type ContractState = {
  last_id: int,
  lockups: int -> Lockup
}
var contract_state: ContractState
\end{lstlisting}

\subsubsection*{Finding actions}
Quint actions define the transitions of the model, and they are the only kind of definition where state variables can be modified. CosmWasm uses values of types \texttt{Deps} and \texttt{DepsMut} to pass the state around, the latter allowing for it to be updated. The tool looks for the presence of those values to determine if a Rust function should be translated into a Quint action or just a pure definition (which is the Quint version of a pure function).

When the tool generates a Quint action, it will also generate a matching pure definition that takes the two parameters in addition to the Rust parameters for the function: \texttt{state: ContractState} and \texttt{deps: Deps} parameter. In Quint, there is no distinction between \texttt{Deps} and \texttt{DepsMut} since there is no mutability.
The pure definition will return the updated value of the contract state. 
Since there is no translation of the body of Rust functions, this will be a stub like the one in Figure~\ref{fig:listings-stubs} in Appendix~\ref{sec:additional-results}.

The generated action itself non-deterministically picks values for all of the Rust function's parameters (except for \texttt{Deps}), builds an execution message with those parameters and calls \texttt{execute\_message} with it.\
\texttt{execute\_message} is a helper action that non-deterministically picks additional metadata (the sender address, funds denomination and funds amount) and calls the \texttt{execute} pure definition, which is a direct translation of the CosmWasm \texttt{execute} function, present in every contract. This is where each message will be processed by the pure definition generated for it (explained in the paragraph above). At each step of the model, \texttt{execute\_message} is called, its result value is assigned to the contract state, and the bank is updated with the funds transfer according to the non-deterministic funds pick.

The bank is a separate state variable that stores the balances for each account. On top of the bank and the contract state, the model also has a variable to track time, and a variable to store the last result--the returned value from the last entry point that was executed which, like in CosmWasm, is either an error or a response object (containing arbitrary attributes and any number of messages).

\subsubsection*{Messages}

Another core feature of CosmWasm's execution environment is how messages are processed. A contract is not able to directly update the bank or call other contracts--it does that through messages. At this point, the mechanical generation tool is able to handle bank messages but not general inter-contract communication. That is an important improvement to be made.

Handling bank messages in the Quint model means that, at every step, if there is a message in the last execution's response, the step should consist of processing that message instead of another execution call. It then updates the bank with the corresponding data inside the message and removes it from the set of messages in the response state variable.

The same concept can be applied for inter-contract communication, as that interaction is also done by including a message in the response. There is additional complexity in processing these messages because it involves other contract's state and types.

\subsubsection*{Generated adapter}

With the goal of bridging the gap between the Quint model and the CosmWasm code, the generator tool produces an adapter file. This is a Rust test that reads JSON traces produced by Quint tooling and reproduces each step of the trace in the CosmWasm contract, using the testing library \texttt{cw\_multi\_test}~\cite{cwMultiTest}, which defines helpers to interact with the contract. If the trace can be reproduced successfully, and the states after each execution match, it is evidence that the set of behaviors defined by Quint model is also defined in the implementation. This test can be run for an arbitrary number of traces---the larger the number of traces we generate, the higher our confidence is that the implementation conforms to the model.

\subsection{LLM-based Synthesis for Contract Logic}
\label{sec:llm-synthesis}

In this phase of the generation, we instruct the LLM to complete the Quint stubs, i.e., the partial model with placeholders in place of functions that implement the contract logic.
In addition to the stubs that resulted from the mechanical generation phase, we assume that the following three types of inputs will be made available by the user.

\begin{enumerate}
    \item CosmWasm source code, which contains the data structures that the contract stores on the blockchain, Rust implementations of message handlers, and helper data structures such as possible error formats.
    \item Natural-language descriptions of \jan{the functionality that each stub is supposed to encode}, which will steer the LLM's output towards Quint code that correctly captures the contract logic.
    \item Two input-output specifications for each placeholder, which will be used to steer generation, and repair the generated code in case of errors. 
\end{enumerate}

The result of the mechanical generation phase is a Quint model with ``stubs'' in place of implementations of pure Quint functions that capture the functionality of the smart contract message handlers.
They correspond to CosmWasm message handlers in the contract's Rust source code (and any utility functions that they may use).
Stubs are \emph{pure} Quint functions that are intended to capture the main contract logic.
Pure functions are defined solely by their input-output behavior and have no side effects, i.e., they do not modify or read any external state implicitly, nor perform any I/O operations.
The core idea of our approach is to factorize the complex task of generating an entire model into a set of more tractable pure function synthesis tasks.
Crucially, the LLM may complete each stub \emph{independently} of others.
This also applies to testing and repairing the generated implementations.

Relying on this factorization, we provide a pseudocode outline for our LLM-based synthesis method in Algorithm~\ref{alg:llm-synthesis}.
This pseudocode is run for each of the stubs.
In Section~\ref{sec:few-shot} \emph{(In-Context Learning)} we explain how the LLM is prompted to generate the code for a particular stub.
We explain the final phase of testing and repairing each generated implementation in Section~\ref{sec:repair}.
Throughout this work, we sample the LLM using a temperature of $0.6$ and a top-p value of $0.7$.
Our implementation of LLM-based synthesis is available on GitHub.\footnote{\url{https://github.com/informalsystems/FuzzMo/}}

\begin{algorithm}
\caption{LLM-Based Synthesis and Repair (Single Stub)}\label{alg:llm-synthesis}
\begin{algorithmic}[1]
\Require \textit{partialModel}, \textit{cosmWasmSource}, \textit{nlDescriptions}, \textit{ioExamples}, \textit{stubName}, \textit{stubDescription}
\State \textit{prompt} $\gets$ \Call{GeneratePrompt}{\textit{stubName}, \textit{stubDescription}, \textit{ioExamples}, \textit{cosmWasmSource}, \textit{partialModel}}
\State \textit{code} $\gets$ \Call{CallLLM}{\textit{prompt}}
\State \textit{stub} $\gets$ \Call{Extract}{\textit{stubName}, \textit{partialModel}}
\State \textit{filledStub} $\gets$ \Call{Replace}{\textit{partialModel}, \textit{stub}, \textit{code}}
\State \Return \newline \Call{TestAndRepair}{\textit{stubName}, \textit{ioExamples}, \textit{filledStub}}
\end{algorithmic}
\end{algorithm}

\subsubsection*{In-Context Learning}
\label{sec:few-shot}
It is well-understood that large language models are capable of in-context learning~\cite{DBLP:journals/corr/abs-2005-14165}, i.e., learning from examples prepended to the main prompt.
Instead of fine-tuning the LLM on a downstream task such as code generation, which is a slow and data-hungry process, one may improve the model's performance by seeding the context window with a number of example tasks, along with solutions that demonstrate the desired output.

This format of providing few-shot examples which facilitate in-context learning is commonly accepted by LLM APIs natively in the form of a messaging API.
Each message in OpenAI's Chat Completions API is tagged as either a \emph{system} message, which is used to provide general instructions or set the tone of the conversation, \emph{user} message which contains the user query in the form of a question, or an \emph{assistant} message, demonstrating the expected answer.

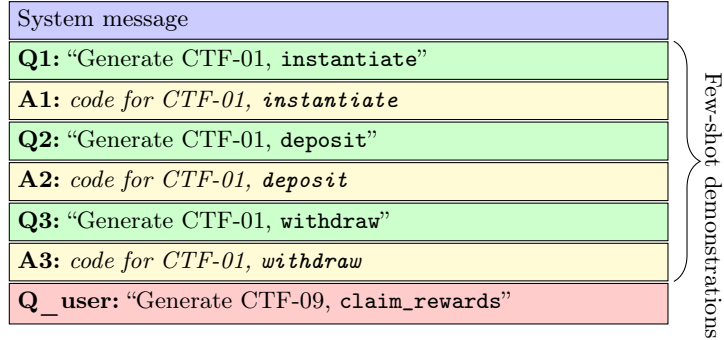
\begin{figure}[H]
    \centering
    \begin{tikzpicture}[
        node distance=0.2mm,
        message/.style={rectangle, draw, align=left, minimum height=5mm, text width=0.7\columnwidth},
        system/.style={fill=blue!20},
        user/.style={fill=green!20},
        assistant/.style={fill=yellow!20},
        final/.style={fill=red!20}
    ]
        \node[message, system] (system) {System message};

        \node[message, user, below=of system] (q1) {\textbf{Q1:} ``Generate CTF-01, \texttt{instantiate}''};
        \node[message, assistant, below=of q1] (a1) {\textbf{A1:} \textit{code for CTF-01, \texttt{instantiate}}};
        
        \node[message, user, below=of a1] (q2) {\textbf{Q2:} ``Generate CTF-01, \texttt{deposit}''};
        \node[message, assistant, below=of q2] (a2) {\textbf{A2:} \textit{code for CTF-01, \texttt{deposit}}};

        \node[message, user, below=of a2] (q3) {\textbf{Q3:} ``Generate CTF-01, \texttt{withdraw}''};
        \node[message, assistant, below=of q3] (a3) {\textbf{A3:} \textit{code for CTF-01, \texttt{withdraw}}};

        \node[message, final, below=of a3] (q_user) {\textbf{Q\_user:} ``Generate CTF-09, \texttt{claim\_rewards}''};

        \draw[decorate,decoration={brace,amplitude=10pt,raise=2pt}] (q1.north east) -- (a3.south east) 
        node[midway,xshift=6mm, yshift=13mm,rotate=-90] {Few-shot demonstrations};
    \end{tikzpicture}
    \caption{Message format illustrating the use of few-shot examples for in-context learning.}
    \label{fig:message-format}
\end{figure}

The final message in the sequence is commonly a user message containing the desired prompt that the LLM is supposed to answer in the style of preceding messages.
Figure~\ref{fig:message-format} illustrates the structure of our in-context learning prompt format when generating based on a stub called \texttt{claim\_rewards} (a part of CTF-09), demonstrating the sequence of system message, few-shot examples, and the final user query.

In our few-shot examples, each question emulates a user asking the model to generate a particular Quint pure function that encapsulates the logic of a message handler found in CTF-01 (\texttt{instantiate}, \texttt{deposit}, or \texttt{withdraw}).
The answer, emulating the LLM's response, is the correct implementation taken from our hand-crafted model for CTF-01.
A partial template we use to create the question prompts is provided in Listing~\ref{lst:partial-prompt-template} (see Appendix~\ref{sec:full-user-template} for the full version).
We use the same template to format the final user prompt, containing the query for the target Quint function of the model we are currently generating.

As shown by our experimental evaluation, presented in part in Table~\ref{tab:ctf-04-09-diff-3}, this prompting style resulted in highly successful initial generations, frequently eliminating the need for further repair steps.
We hypothesize that this high success rate is due to two key factors.
First, the mechanical generation phase manages a significant portion of the task, reducing the potential for errors that would otherwise burden the LLM.
Second, we enrich the LLM's context window with highly relevant information derived from analyzing generated stubs and the Rust source code of the contract.

\begin{lstlisting}[language=, style=templatestyle, caption={Partial user prompt template},label=lst:partial-prompt-template]
Please complete this stub for my `@@@NAME@@@` Quint function.
@@@DESCRIPTION@@@
```
@@@STUB@@@
```
Here is the contract state type and related types:
```
@@@QUINT TYPE DEFINITIONS@@@
```
<omitted>
The `@@@NAME@@@` Quint function models a CosmWasm Rust function for processing smart contract messages.
@@@MESSAGE HANDLER@@@
Your Quint code must be functional.
All function arguments are immutable, and you can not use mutable variables.
@@@IO EXAMPLES@@@
\end{lstlisting}

In the template from Listing~\ref{lst:partial-prompt-template}, the \texttt{\textcolor{templatekeyword}{@@@NAME@@@}} macro expands to the stub (pure function) name, such as \texttt{withdraw}, as demonstrated in the third few-shot example for CTF-01 in Figure~\ref{fig:message-format}.
The \texttt{\textcolor{templatekeyword}{@@@DESCRIPTION@@@}} macro expands to the user-provided natural language description of the stub's intended functionality, as illustrated in Listing~\ref{lst:withdraw-description} for the \texttt{withdraw} stub in CTF-01.
We obtain the expansion for \texttt{\textcolor{templatekeyword}{@@@QUINT TYPE DEFINITIONS@@@}} by analyzing the output of the mechanical generation phase.
Starting from the function signature of the placeholder at hand, we recursively look up relevant type definitions in the stub model's Quint source code.

Coupled with a Quint cheatsheet with a list of commonly used operations that we always include in the system message, this provides the LLM with useful and targeted knowledge to execute complex operations on data and implement the desired functionality for a particular stub.

In order to expand the \texttt{\textcolor{templatekeyword}{@@@MESSAGE HANDLER@@@}} macro, we search the contract's CosmWasm source code for a block of Rust code that is a function definition beginning with \texttt{pub fn <stub name>}.
Our mechanical generation tool guarantees that this function (the message handler which induced the placeholder) exists in the Rust source files.

Finally, the \texttt{\textcolor{templatekeyword}{@@@IO EXAMPLES@@@}} macro expands to the input-output specifications provided by the user.
(In our evaluation, we supply the LLM with two input-output specifications.)
The full version of this template can be seen in Listing~\ref{lst:full-user-template} in Appendix~\ref{sec:prompt-templates}.


\begin{lstlisting}[language=, style=descriptionstyle, caption={User-provided description for CTF-01, withdraw stub},label=lst:withdraw-description]
"Enable users to withdraw from valid lockups, updating the state, or return an error if validation fails."
\end{lstlisting}

\subsubsection*{Adapter Generation for Model-based Testing}
\label{sec:adapter-generation}

As explained in Section~\ref{subsec:mbt}, model-based testing involves executing traces, sequences of state transitions generated by the model, against the implementation.
An adapter is a Rust test which accomplishes this, but writing one manually is a time-consuming task.
In order to fully automate a model-based testing workflow, we generate an adapter by following a procedure similar to the one outlined throughout Section~\ref{sec:model-generation}: a baseline test which executes the trace can be generated mechanically, but we use an LLM to synthesize the state comparison function, a part of the adapter which checks whether contract states correspond with state transitions in the model.

While a trivial state comparison function can be mechanically generated (e.g., one that always returns true), we leverage the LLM's instruction-following capabilities to generate a function that performs a more thorough inspection of the smart contract's state.
If we detect errors in the generated adapter code, we trigger a repair loop similar to the one we outline for model generation in the next section.
In Appendix~\ref{sec:case-study-mbt}, we conduct a case study for the model-based testing workflow supported by adapter generation, and relegate the technical details to Appendix~\ref{sec:adapter-details}.

\subsection{Testing and Repair Phase}
\label{sec:repair}

In this section, we outline our strategy for addressing potential errors that arise during generation.
As highlighted in Section~\ref{sec:llm-synthesis}, we independently repair each function in the model.
Because we never generate code with side effects, it suffices that the generated code correctly implements the intended behavior for all I/O specifications.
This stems from the fact that the model's overall structure and state management code have been generated deterministically.
We divide errors into three distinct categories, based on the earliest point at which they can be detected, namely
\begin{enumerate}
    \item \textbf{static errors} (e.g., incorrect Quint syntax), which are detected before executing any tests;
    \item \textbf{runtime errors} (e.g., calling a nonexistent method), which are detected when the generated is executed; and
    \item \textbf{semantic errors}, which are detected once the generated pure function returns an output that is inconsistent with a user-provided input-output specification.
\end{enumerate}

After the initial generation is completed, we analyze it for the presence of either static, runtime, or semantic errors.
If we detect them, our strategy is to enter a loop of test-and-repair rounds, as shown in Figure~\ref{fig:method}.
In each round, we gather relevant information about the probable cause (such as error messages) and construct a prompt asking the LLM to repair the code based on this context.
The LLM's response then replaces the previous, erroneous fragment of code, and we check the new version for correctness again.

In order to prevent an unbounded number of LLM calls, we define a finite repair round ``budget'' for each error category: the amount of times the system may invoke an LLM to repair a particular type of error.
In all of our experiments, we use a budget of $3$ repair rounds for each error category, leading to a maximum of $9$ additional LLM calls.

Although this approach does not guarantee an error-free result, in practice we have found it sufficiently capable of achieving our goal: reducing the developer time necessary to utilize Quint models.
As shown in our experimental evaluation, in most runs we were able to successfully generate correct code for most target functions and contracts in our benchmark, using fewer than $3$ rounds of repair on average.
We outline the pseudocode for this phase (the $\textsc{TestAndRepair}$ function in Algorithm~\ref{alg:llm-synthesis}), in Algorithm~\ref{alg:llm-repair} of Appendix~\ref{sec:llm-repair}.

In the following two sections, we further explain the static and semantic repair rounds.
When repairing runtime errors, we use 
\jan{a combination of}
methods and prompts employed for static and semantic errors (details in Appendix~\ref{sec:runtime-errors}).

\subsubsection*{Static Errors}
\label{sec:static-errors}
To detect static errors, we utilize Quint's type checking tool (which will, naturally, also detect syntax errors).
An example of this tool's output when detecting a syntax error is presented in Listing~\ref{lst:static-error}.
This error occurred during code generation for a stub named \texttt{execute\allowbreak\_add\allowbreak\_strategy} in our real-world Neutron DAO case study.

Quint being a new language means that commercial LLMs lack comprehensive information on Quint-specifics.
The error in Listing~\ref{lst:static-error} arises from the new and limited pattern matching syntax in Quint.
The generated code attempts to pattern-match on literal values, a valid pattern in many languages that support this feature.
However, as the error message points out, this is not yet supported in Quint.
(In this case, a parser-informed method such as PICARD~\cite{DBLP:journals/corr/abs-2109-05093} would prove valuable.)

Using our repair prompting methodology, the LLM is able to correctly reason about the root cause of the error, and subsequently produces correct Quint code.
Although we lack a dataset of few-shot examples suitable for our repair phase, we compensate by instructing the LLM to start the solution with reasoning about the problem before writing any code \cite{DBLP:journals/corr/abs-2201-11903}.
For brevity, we omit the corrected code, and instead showcase the LLM's reasoning in Listing~\ref{lst:static-error-fixed}.
The template we use to generate static repair prompts can be found in~Appendix~\ref{sec:full-repair-user-template}.

\begin{lstlisting}[language=, style=errorstyle, caption={Static error in Neutron DAO},label=lst:static-error]
[...]quint/model_generated.qnt:86:20 - error: [QNT000] mismatched input 'true' expecting {'_', LOW_ID, CAP_ID}
86:    | Ok(true) => (Err("Cannot remove last administrator"), state)
            ^^^^

[...]quint/model_generated.qnt:87:20 - error: [QNT000] mismatched input 'false' expecting {'_', LOW_ID, CAP_ID}
87:    | Ok(false) => (
            ^^^^^
\end{lstlisting}

\begin{lstlisting}[language=, style=descriptionstyle, caption={LLM reasoning},label=lst:static-error-fixed]
2. **Mismatched input 'true' expecting {'_', LOW_ID, CAP_ID}**: This error indicates that the use of the boolean literals `true` and `false` directly in pattern matching is not allowed in Quint.
3. **Mismatched input 'false' expecting {'_', LOW_ID, CAP_ID}**: Similar to the previous error, the use of the boolean literal `false` directly in pattern matching is not allowed.
\end{lstlisting}

\subsubsection*{Semantic Errors}
\label{sec:semantic-errors}

After embedding the synthesized function into the mechanically generated stub file and confirming it passes type checking, we verify whether it respects the user-provided I/O specifications.

Let $F$ denote the true Quint pure function that implements a particular part of the contract's logic in the model.
In this setting, $F$ is unknown, and we only assume access to $K \in \naturals$ user-provided I/O specifications ($K \geq 2$), and $\tilde{F}$, the LLM's completion for a stub of $F$.
Mechanical generation guarantees that $F$ and $\tilde{F}$ share the same type signature (arguments and return type), and we require that $\tilde{F}$ must pass type checking before semantic repair rounds are triggered.
Thus, we may call $\tilde{F}$ with arguments suitable for $F$, and if $\tilde{F}$ returns, the return value will be comparable with the return value of $F$.

Semantic repair rounds run only if there are no runtime errors.
Each I/O specification is represented as a pair $(\text{args}_i, R_i)$ ($1 \leq i \leq K$), 
where $\text{args}_i$ denotes the arguments applied to $F$ or $\tilde{F}$ (the user-specified input), 
and $R_i=F(\text{args}_i)$ denotes the expected return value (the user-specified output).
We identify a semantic error if $\tilde{F}(\text{args}_i) \neq R_i$ for a given $i$.
In that case, we start a semantic repair round by preparing a specialized prompt, employing a template similar to the one used for static errors.

We provide to the LLM (1) $\text{args}_i$, the user-specified input data; (2) $\tilde{F}(\text{args}_i)$, the incorrect output of the current implementation; and (3) $R_i$, the user-specified expected output.
Of course, just as in a static repair round, the prompt also contains other information such as the erroneous implementation ($\tilde{F}$).
In this way, we provide sufficient information to correct the function's behavior at a number of points in the input space, and we rely on the LLM's generalization capabilities to enhance the overall correctness.

\section{Evaluation}
\label{sec:evaluation}

In Section~\ref{sec:ctf-benchmark}, we evaluate our Quint model generation technique on a benchmark consisting of $5$ smart contracts from the CosmWasm Capture the Flag repository and present summary results.
In Appendix~\ref{sec:additional-results}, we present more comprehensive results for the entire benchmark, and perform a limited ablation study investigating the effect of natural language descriptions on performance across the CTF Benchmark.

In Section~\ref{sec:case-study}, we present a case study applying our method to a real-world audit scenario.
An additional case study, capturing the full model-based testing workflow, can be found in Appendix~\ref{sec:case-study-mbt}.

\subsection{Benchmark: CosmWasm Capture the Flag}
\label{sec:ctf-benchmark}

The CosmWasm Capture the Flag (CTF) repository provides the source code for $10$ smart contracts used in a capture-the-flag competition.
Each smart contract in the repository contains a critical bug, and the goal in a CTF competition is to find the bug and craft an exploit around it.
For example, in CTF-01, users may deposit funds into a lockup, and withdraw funds from lockups by specifying a lockup ID.
The bug can be found in the \texttt{withdraw} message handler, which does not check if the same lockup ID appears multiple times in the message, enabling users to withdraw all funds from the same lockup twice, and thus steal funds from the contract.
We focused our benchmark on contracts from the CosmWasm CTF repository where the vulnerability is caused by a bug in program logic rather than by a technical or platform-dependent reason, making Quint models useful for detecting the issue.

Our benchmark consists of $23$ target functions, drawn from our hand-crafted reference Quint models for CTFs 02, 04, 05, 07, and 09.
We used the CTF-01 \jan{contract} and its model for in-context learning, as described in Section~\ref{sec:few-shot}: we crafted a series
of Question-Answer pairs, each pair demonstrating how to
fulfill the task of generating implementations (Answer) given
Quint placeholders (Question).

Furthermore, we wrote a natural-language description of each target function, along with $5$ I/O examples.
We use $2$ I/O examples to emulate a user's input, and the remaining $3$ as a partial specification for post-generation inspection.
(Each function was additionally manually inspected for correctness.)

In this benchmark, our tool generated and repaired a Quint model for each of the contracts.
To manage stochasticity inherent in LLM outputs, we repeat this experiment across $5$ independent runs.
We made a best-effort attempt to maximize the reproducibility of our experiments.
In addition to a frozen model checkpoint, we provide timestamps, system fingerprints, and the seeds used for all LLM API calls.

We present the results of our benchmark for CTF-04 and 09 in Table~\ref{tab:ctf-04-09-diff-3}, showing that the final, post-repair, results generate a function that satisfies the supplied I/O specification in almost all runs (indicated by the green checkmark in the first column).
Full results can be found in Appendix~\ref{sec:additional-results}, in Table~\ref{tab:ctf-difficulty-3-study-3}.

We supply the first two I/O examples as part of the initial generation request, and to perform test-and-repair rounds.
The column \textbf{Avg. Repair} indicates the average number of repair rounds (static, runtime, and semantic) that were used for generating each function.
If the generated code contains no runtime errors, we evaluate the functions on the remaining three I/O examples, that the LLM did not have access to during generation, and report the results as the \textbf{Pass Rate}.





\begin{table}[h]
    \centering
    \caption{Average number of repair rounds used per function and pass rates for unseen test cases, averaged across $5$ independent runs for CTF-04 and CTF-09}
    \begin{tabular}{
        |>{\centering\arraybackslash}p{1.9cm}
        >{\centering\arraybackslash}p{3cm}
        >{\centering\arraybackslash}p{3cm}
        >{\centering\arraybackslash}p{3cm}|}
    \hline
    \multicolumn{4}{|c|}{\underline{\textbf{CTF-04}}} \\
    \multicolumn{4}{|c|}{} \\[-0.8em]
    & \textbf{Function} & \textbf{Avg. Repair} & \textbf{Pass Rate} \\ \hline
    \cmark & \texttt{instantiate} & 0.0 & 100\% \\
    \cmark & \texttt{mint} & 1.2 & 90\% \\
    \cmark & \texttt{burn} & 0.0 & 100\% \\ \hline
    \end{tabular}
    
    \begin{tabular}{
        |>{\centering\arraybackslash}p{1.9cm}
        >{\centering\arraybackslash}p{3cm}
        >{\centering\arraybackslash}p{3cm}
        >{\centering\arraybackslash}p{3cm}|}
    \hline
    \multicolumn{4}{|c|}{\underline{\textbf{CTF-09}}} \\
    \multicolumn{4}{|c|}{} \\[-0.8em]
    & \textbf{Function} & \textbf{Avg. Repair} & \textbf{Pass Rate} \\ \hline
    \cmark & \texttt{deposit} & 2.4 & 73\% \\
    \mixed \; (4/5) & \texttt{withdraw} & 1.6 & 73\% \\
    \cmark & \texttt{increase\_reward} & 0.2 & 100\% \\
    \mixed \; (4/5) & \texttt{claim\_rewards} & 1.8 & 80\% \\ \hline
    \end{tabular}
    \label{tab:ctf-04-09-diff-3}
\end{table}

To provide a clearer view of each contract's difficulty, Table~\ref{tab:repair-rounds} details the total number of repair rounds performed per contract, categorized by type of repair.

\begin{table}[h]
    \centering
    \caption{Distribution of error repair rounds by type}

    \begin{tabular}{|c|c|c|c|c|}
        \hline
        \textbf{Contract} & \textbf{Static} & \textbf{Runtime} & \textbf{Semantic} & \(\boldsymbol{\Sigma}\) \\
        \hline
        CTF-02 & 0 & 0 & 3 & 3 \\
        \textbf{CTF-04} & \textbf{1} & \textbf{5} & \textbf{0} & \textbf{6} \\
        CTF-05 & 0 & 0 & 0 & 0 \\
        CTF-07 & 4 & 0 & 0 & 4 \\
        \textbf{CTF-09} & \textbf{23} & \textbf{0} & \textbf{7} & \textbf{30} \\
        \hline
    \end{tabular}
    \label{tab:repair-rounds}
\end{table}

More detailed results can be found in Appendix~\ref{sec:additional-results}. They follow the trends described in this section and provide one additional insight: the quality of the NL description degrades the performance only slightly, allowing the method to work well even if the user fails to provide a precise description.

\subsection{Case Study: Privileged subDAO Contract}
\label{sec:case-study}
In this case study, we applied the described approach to a project from one of the audits that Informal performed recently.
The scope of the audit was a new functionality of Neutron's~\cite{neutron} governance contracts.
To provide sufficient context, let us describe the basic concepts of on-chain governance.

Most blockchains, including Neutron, typically do not have a single chain owner.
Instead, the chain is jointly owned and governed by all the token holders.
Any user may create a \emph{governance proposal}, and the token holders vote for or against it.
If there is sufficient support for the proposal, it gets automatically executed on the chain.
We call such an organizational structure a \emph{decentralized autonomous organization} (DAO).

The whole governance process, from proposal creation to its execution, takes a non-negligible amount of time.
Furthermore, some proposals may be very technically specialized (e.g., regular adjustment of particular chain parameters) and thus not sufficiently interesting for members of the DAO.
To address these issues, the newly added functionality introduced the concept of a \emph{privileged subDAO}: a smaller set of users authorized to vote in and execute specific changes only (before execution, the decision can still be overridden by members of the main DAO).

The functionality is implemented in the \texttt{neutron-chain-manager} contract.
Different subDAOs can either have an \texttt{AllowAll} authorization (typically, for the main or security DAO, which can do anything) or an \texttt{AllowOnly([Permission])} authorization, which specifies a set of permissions that a subDAO can execute.
The relevant entry point functions of the contract are \texttt{AddStrategy} (which gives an authorization to a subDAO), \texttt{RemoveStrategy} (which removes it), and \texttt{ExecuteMessages} (which executes a message if it has authorization to do so). Overall, the contract contains 16 functions that together implement the desired functionality.

The main property that needs to be preserved is the following: \texttt{ExecuteMessages} that attempts to execute a message for which the sender subDAO does not have permission will fail. 
Unfortunately, in the audited commit, the property was not satisfied. 
The problem was identified through manual code analysis and fixed for the release.

\subsubsection*{Identifying the bug using model generation}
We took the code at the audit commit and wanted to see if we could find the same bug using model generation.
We first applied the mechanical generation step described in Section~\ref{sec:mechanical-generation}, and then the model generation described in Section~\ref{sec:llm-synthesis}.
The artifacts resulting from this case study are available on GitHub.\footnote{\url{https://github.com/informalsystems/fuzzmo_use_case2/}}

Mechanical generation reliably produced stubs as expected. 
There was some additional cleanup work needed: removing references to external libraries or unused types, and removing stubs for contracts that were out of scope.
Furthermore, some adjustment was needed for handling message passing of the custom \texttt{NeutronMsg} type. 
Overall, the effort to go from the generated version to a working model stub was neither trivial nor too big.

Out of 16 functions that there were in the contract, we identified 6 of them relevant for deciding on the correct authorization.
For those 6 functions, we ran the model generation process.

Out of 6 functions, 4 were immediately generated correctly. 
The remaining two functions were fixed with syntactic repair. 
The problem was that at the first attempt, there were issues in calling functions within the generated code (once the name was hallucinated, and once the order of parameters did not match the signature).

One interesting observation compared to the reference model (a model written by the authors manually): at a couple of places we had to make sure that a property holds for all elements of a list. 
Thus, we factored that functionality out and wrote the helper function \texttt{listforall} (following the naming of the equivalent function on sets, \texttt{forall}).
The generated model, since it always generates only one function at a time (without the possibility of creating helper functions), generated expressions that were more complicated compared to the reference implementation.

Finally, the generated Quint model sufficed to uncover the authorization bug that existed in the contract.
This presents the main takeaway of the case study: the presented approach could indeed be used in a real-world audit.

Beside the main takeaway, there are a couple of peculiarities worth commenting on.
First, we see that there was no need for semantic repair of generated model functions.
This is because the authorization mostly depends on getting the types right, and the logic is fairly simple.

Second, because our model stub ensures full generality handling CosmWasm bank transfers, and the analyzed contracts did not include any, the whole model ended up less readable than a model an experienced Quint user would have written by hand.
The tradeoff is then between ease of modeling and ease of reading.
Because these models are used for bug finding, and not for protocol documentation (which is another use case for Quint models), we find that making modeling easier is the right choice for all but the most proficient Quint modelers.
Contrary to the full generality of mechanically generated stubs, we found the LLM-generated parts of the models to be written in a very natural way.

There are two caveats regarding the optimistic takeaways of this section.
First, our method presents the core around which an ergonomic, 
user-friendly tool should be built.
Second, generalizing the method to smart contract environments other than CosmWasm would require separate design process for the stub generation part.

\section{Conclusion}
\label{sec:conclusion}

In this paper, we reported on our effort to bridge the gap between model-based techniques and software auditing practice.
We did so by automatically generating both Quint models and model-based tests, taking a CosmWasm contract, I/O examples, and a high-level description as input.
We demonstrated how our technique is able to generate correct models on a capture-the-flag benchmark, and ran a successful case study on a past audit.

Our experience shows that LLMs can be tamed by using mechanical generation for setting up a firm frame for statistical code generation.
Iterated repair attempts (syntactic, semantic, and runtime) were useful for correcting the LLM's occasional hallucinations.
While the proposed method does not guarantee correctness, our evaluation shows that it works well in practice.
Furthermore, at every step of the generation process (model stubs, models, test stubs, tests), the method provides useful artifacts for the users, which is very much appreciated by practitioners.

There are still aspects of our technique that need to be improved if we want to make it fully practical.
Currently, the input-output examples we expect from users are encoded as Quint files with Quint syntax.
Providing a better user experience there would go a long way to bring our technique to a broader audience.
Additionally, mechanical generation, while fully reliable, is still restricted to CW contracts that do not send messages among themselves.
This limits the applicability of the technique to simpler contracts.

The main practical benefit of our approach is easing auditors' way into using model-based techniques---a task with a very steep learning curve prior to having our tool available.

\clearpage

%
%
%
\bibliographystyle{splncs04}
\bibliography{references}

\clearpage

\appendix

\section{Case Study: Model-based Testing Workflow}
\label{sec:case-study-mbt}

We conducted a case study of the model-based testing workflow---which includes both model generation and adapter generation---on two smart contracts, CTF-02 and 04.
First, we simulated Quint models generated as part of the benchmark from Section~\ref{sec:ctf-benchmark} to obtain traces.
We then generated adapters that can execute these traces, as discussed in Section~\ref{sec:adapter-generation}.
In Appendix~\ref{sec:adapter-details} we provide further details on the adapter generation process in this case study, including examples of state comparison functions synthesized with our LLM-based method.

Listing~\ref{lst:adapter-output} shows the output produced by our synthesized adapter for CTF-02 while processing a trace generated by our synthesized model of the contract.
As can be seen, the test compiles and successfully starts executing the trace, but fails after two steps, uncovering a bug.
The test fails due to a withdraw message in the trace sent from an unrecognized address.
The Quint model is able to handle this situation gracefully, while the CosmWasm implementation panics.
This difference between the model and the implementation points to a vulnerability in the code.

\begin{lstlisting}[language=, style=errorstyle, caption={Adapter output (CTF-02)},label=lst:adapter-output]
Step number: Some(0)
Result from trace: Ok(Response { messages: [], attributes: [Attribute { key: "action", value: FromStr("instantiate") }] })
Initializing contract.
Message: InstantiateMsg
Sender: Addr("admin")
Funds: []
Contract balance (Addr("contract0")) for uawesome: Uint128(200) vs Uint128(200)
clock is advancing for 1 seconds
-----------------------------------
Step number: Some(1)
Result from trace: Err("Unauthorized (Withdraw)")
Message: Withdraw { amount: Uint128(32) }
Sender: Addr("contract0")
Funds: [Coin { 77 "d1" }]
thread 'tests::model_test' panicked at src/contract.rs:68:66:
called `Result::unwrap()` on an `Err` value: NotFound { kind: "type: oaksecurity_cosmwasm_ctf_02::state::UserInfo; key: [00, 0C, 76, 6F, 74, 69, 6E, 67, 5F, 70, 6F, 77, 65, 72, 63, 6F, 6E, 74, 72, 61, 63, 74, 30]" }
\end{lstlisting}

Each CTF contract is designed to demonstrate only a specific exploit, and they do not implement handling of various edge-cases such as this.
Thus, this was an easy catch for the model-based testing workflow we envisioned, which replaces hours of careful code inspection.
By iteratively fixing all the differences between the code and model, we can eliminate all the bugs in the contract, including the main vulnerability.

Another approach to achieving the same goal (uncovering the main vulnerability) is to modify the adapter so as to catch the panics raised by the contract under test, verify that they correspond to the errors predicted by the model, and if so, ignore them and carry on with processing the trace.
Although this approach is less natural than correcting the model itself, we chose to follow it so as to minimize changes to our dataset (the CTF benchmark).
We cover this process in the following section.

\subsection{Uncovering the Main Vulnerability in CTF-02}
\label{sec:ctf-02-vuln}

The exploit in CTF-02 is triggered when an attacker attempts to unstake (reclaim) more funds than they had originally staked.
In a testing environment, an overflow error will occur in this scenario, reverting the transaction.
This may give developers false confidence that the exploit will similarly be prevented when executing on-chain.
However, the core idea of the vulnerability is that overflow checks will be disabled in the release version of the contract, making it easier for developers (who primarily work within the testing environment) to overlook this bug.

In order to uncover this vulnerability, we needed to accommodate the generated adapter from the previous section to be more resilient against contract panics, such as the one from Listing~\ref{lst:adapter-output}.
Specifically, we modified the adapter to unwind particular panics raised by CTF-02 (e.g., if an unrecognized token denomination is received in a deposit message).
In that case, we ensured that corresponding transitions in the trace also report an error of the same kind in the model (otherwise, we would not be alerted to a worrying mismatch between the model and the implementation).

These modifications allowed the adapter to progress further and process more steps of the trace, up until discovering the main vulnerability.
As can be seen in Listing~\ref{lst:adapter-output-ctf-02-vuln}, the adapter processed $14$ steps of the trace and eventually encountered an inconsistent state, thus uncovering the vulnerability.

\begin{lstlisting}[language=, style=errorstyle, caption={Adapter output (CTF-04)},label=lst:adapter-output-ctf-02-vuln]
Step number: Some(14)
Result from trace: Err("Insufficient voting power (Unstake)")
Message: Unstake { unlock_amount: 194 }
Sender: Addr("sender3")
Funds: [Coin { 29 "uawesome" }]
thread 'tests::model_test' panicked at tests/contract_model_test_adapter_study.rs:172:13:
Expected action to fail with error: Some("Insufficient voting power (Unstake)")
\end{lstlisting}

In order to generate this output, one needs to run \text{cargo test} using the \verb|--release| flag.
The vulnerability is uncovered by our generated adapter, which expects the unstake action to \emph{fail}, while the contract's CosmWasm implementation simply carries on with an overflow in its arithmetic (due to the release flag).
If one leaves out the release flag, our generated adapter will still discover the bug, but the output will be slightly different: instead of complaining about the action succeeding, when it should have failed, this time the implementation itself will panic because the overflow checks are present, successfully alerting the developer of the vulnerability.

\subsection{Uncovering an Unexpected Vulnerability CTF-04}

Similar to our end-to-end model-based testing case study for CTF-02, when executing a trace from the Quint model for CTF-04, our generated adapter panics and inadvertently uncovers a bug in CTF-04.
The output of this test run in shown in Listing~\ref{lst:adapter-output-ctf-04}.
In this case, the problematic state transition in the trace involves an unrecognized token denomination.
The bug is found in the \texttt{mint} function of CTF-04's Rust implementation, where the CosmWasm utility function \texttt{must\_pay} is called without proper error handling, resulting in a panic.~\footnote{\url{https://github.com/oak-security/cosmwasm-ctf/blob/c32fa94947f4199596d26696afeb3de0ab3d9cca/ctf-04/src/contract.rs\#L45}}

\begin{lstlisting}[language=, style=errorstyle, caption={Adapter output (CTF-04)},label=lst:adapter-output-ctf-04]
Step number: Some(0)
Result from trace: Ok(Response { messages: [], attributes: [] })
Initializing contract.
Message: InstantiateMsg { offset: 0 }
Sender: Addr("admin")
Funds: []
Contract balance (Addr("contract0")) for uawesome: Uint128(200) vs Uint128(200)
Contract config total_supply: Uint128(0) vs Uint128(0)
clock is advancing for 1 seconds
-----------------------------------
Step number: Some(1)
Result from trace: Ok(Response { messages: [], attributes: [] })
Message: Mint
Sender: Addr("sender1")
Funds: [Coin { 31 "d1" }]
thread 'tests::model_test' panicked at src/contract.rs:45:41:
called `Result::unwrap()` on an `Err` value: MissingDenom("uawesome")
\end{lstlisting}

\section{Testing and Repair Phase Details}
\label{sec:repair-details}
In this section, we provide additional details on the testing and repair phase of our method that were omitted from the main paper.
The pseudocode for the testing and repair phase can be found in the following Section~\ref{sec:llm-repair} as Algorithm~\ref{alg:llm-repair}.
In Section~\ref{sec:runtime-errors} we describe runtime repair rounds.

\subsection{Full Repair Algorithm Pseudocode}
\label{sec:llm-repair}

We assume that $\textsc{GeneratePrompt}$ calls in Algorithm~\ref{alg:llm-repair} have access to similar contextual information (such as the stub name, description, and CosmWasm source code) as in Algorithm~\ref{alg:llm-synthesis}, but for the sake of brevity we do not explicitly state them as arguments.
The full template used by $\textsc{GeneratePrompt}$ calls in this case can be found in Appendix~\ref{sec:full-repair-user-template}.

\begin{algorithm}
\caption{Testing and Repair Phase}\label{alg:llm-repair}
\begin{algorithmic}[1]
\Require \textit{stubName}, \textit{ioExamples}, \textit{filledStub}, \textit{staticRounds}, \textit{runtimeRounds}, \textit{semanticRounds}
\State \textit{done} $\gets$ \textbf{false}
\While{not \textit{done}}
    \State \textit{done} $\gets$ \textbf{true}
    \If{not \Call{TypeChecks}{\textit{filledStub}} \textbf{and} \textit{staticRounds} > 0}
        \State \textit{done} $\gets$ \textbf{false}
        \State \textit{staticRounds} $\gets$ \textit{staticRounds} - 1
        \State \textit{typeCheckOutput} $\gets$ \Call{QuintTypeCheck}{\textit{filledStub}}
        \State \textit{prompt} $\gets$ \Call{GeneratePrompt}{\textit{filledStub, \textit{typeCheckOutput}}}
        \State \textit{code} $\gets$ \textsc{CallLLM}(\textit{prompt})
        \State \textit{filledStub} $\gets$ \textsc{Replace}(\textit{quintStubs}, \textit{stub}, \textit{code})
        \State \textbf{continue}
    \EndIf
    \If{\Call{TypeChecks}{\textit{filledStub}} \textbf{and} \Call{Crashes}{\textit{stubName}, \textit{ioExamples}, \textit{filledStub}} \textbf{and} \textit{runtimeRounds} > 0}
        \State \textit{done} $\gets$ \textbf{false}
        \State \textit{runtimeRounds} $\gets$ \textit{runtimeRounds} - 1
        \State \textit{runtimeOutput} $\gets$ \Call{RuntimeCheck}{\textit{stubName}, \textit{ioExamples}, \textit{filledStub}}
        \State \textit{prompt} $\gets$ \Call{GeneratePrompt}{\textit{filledStub, \textit{runtimeOutput}}}
        \State \textit{code} $\gets$ \textsc{CallLLM}(\textit{prompt})
        \State \textit{filledStub} $\gets$ \textsc{Replace}(\textit{quintStubs}, \textit{stub}, \textit{code})
        \State \textbf{continue}
    \EndIf
    \If{not \Call{Crashes}{\textit{stubName}, \textit{ioExamples}, \textit{filledStub}}}
        \State \textit{ioIncorrect} $\gets$ \{\}
        \For{(\textit{args}, \textit{expect}) in \textit{ioExamples}}
            \State \textit{output} $\gets$ \Call{executeQuint}{\textit{stubName}, \textit{args}, \textit{filledStub}}
            \If{\textit{output} != \textit{expect}}
                \State \textit{ioIncorrect}[(\textit{args}, \textit{expect})] $\gets$ \textit{output}
            \EndIf
        \EndFor
        \If{size(\textit{ioIncorrect}) > 0 \textbf{and} \textit{semanticRounds} > 0}
            \State \textit{done} $\gets$ \textbf{false}
            \State \textit{semanticRounds} $\gets$ \textit{semanticRounds} - 1
            \State \textit{prompt} $\gets$ \Call{GeneratePrompt}{\textit{ioIncorrect}}
            \State \textit{code} $\gets$ \textsc{CallLLM}(\textit{prompt})
            \State \textit{filledStub} $\gets$ \textsc{Replace}(\textit{quintStubs}, \textit{stub}, \textit{code})
        \EndIf
    \EndIf
\EndWhile
\end{algorithmic}
\end{algorithm}

\subsection{Runtime Errors}
\label{sec:runtime-errors}

In this section we will use the same notation as in Section~\ref{sec:semantic-errors}, where we explain semantic errors: $\tilde{F}$ denotes the LLM-synthesized function, and $(\text{args}_i, R_i)$ ($1 \leq i \leq K$) denote the $K$ user-provided I/O specifications ($\text{args}_i$ and $R_i$ are the user-specified input and output, respectively).

Runtime errors occur when the synthesized function $\tilde{F}$ attempts to execute illegal Quint code which was not detected by the Quint type checker (such as attempting to call a nonexistent method), or when certain runtime invariants are broken, for example if attempting to access a list with an invalid index.
We detect this by passing all the user-provided arguments $\text{args}_i$, one by one, to $\tilde{F}$, as soon as it passes type checking.
If the execution of this function crashes instead of returning a value, we trigger a runtime repair round.

An example of a runtime error and the synthesized Quint function that caused it are provided in Listings \ref{lst:runtime-error} and \ref{lst:runtime-error-code}, respectively.
As can be seen, this runtime error was caused by map lookup with a missing key.
Finally, we provide the LLM's reasoning based on the error message from Listing~\ref{lst:runtime-error}, including the repaired function, in Listing~\ref{lst:runtime-error-fixed}.

\begin{lstlisting}[language=, style=errorstyle, caption={Runtime error when running CTF-04, burn function},label=lst:runtime-error]
runtime error: error: [QNT507] Called 'get' with a non-existing key
 val user_balance = state.balances.get(info.sender).amount
                    ^^^^^^^^^^^^^^^^^^^^^^^^^^^^^^^
\end{lstlisting}

\begin{lstlisting}[language=quint, label=lst:runtime-error-code, style=quintstyle, caption={CTF-04, burn (erroneous)}]
pure def burn(state: ContractState, deps: Deps, env: Env, info: MessageInfo, shares: int): (Result[Response, ContractError], ContractState) = {
    val contract_balance = deps.querier.bank.get(CONTRACT_ADDRESS).get(DENOM)
    val total_assets = contract_balance
    val total_supply = state.config.total_supply

    // Compute the amount of assets to return
    val asset_to_return = shares * total_assets / total_supply

    if (asset_to_return == 0) {
      (Err("ZeroAmountNotAllowed"), state)
    } else {
      val user_balance = state.balances.get(info.sender).amount

      if (user_balance < shares) {
        (Err("unsigned int error"), state)
      } else {
        val updated_user_balance = user_balance - shares
        val updated_total_supply = total_supply - shares

        val updated_balances = state.balances.put(info.sender, { amount: updated_user_balance })
        val updated_state = { ...state, config: { total_supply: updated_total_supply }, balances: updated_balances }

        val msg = CosmosMsg_Bank(BankMsg_Send({ to_address: info.sender, amount: [{ denom: DENOM, amount: asset_to_return }] }))

        (Ok(Response_new
          .add_attribute("action", FromStr("burn"))
          .add_attribute("user", FromStr(info.sender))
          .add_attribute("asset", FromInt(asset_to_return))
          .add_attribute("shares", FromInt(shares))
          .add_message(msg)), updated_state)
      }
    }
  }
\end{lstlisting}

\begin{lstlisting}[language=, style=descriptionstyle, caption={LLM reasoning for runtime error in CTF-04, burn},label=lst:runtime-error-fixed]
### Explanation of Errors

- **Non-existing key error**: The error `Called 'get' with a non-existing key` indicates that the key `info.sender` does not exist in the `state.balances` map. This means that the code attempts to access a balance for a user that might not have an entry in the `balances` map.

### Repaired Function

To fix this issue, we need to handle the case where `info.sender` does not have an existing balance entry. We can use the `getOrElse` function to provide a default balance of 0 if the user does not have an entry in the map.

```quint
pure def burn(state: ContractState, deps: Deps, env: Env, info: MessageInfo, shares: int): (Result[Response, ContractError], ContractState) = {
    val contract_balance = deps.querier.bank.get(CONTRACT_ADDRESS).get(DENOM)
    val total_assets = contract_balance
    val total_supply = state.config.total_supply

    // Compute the amount of assets to return
    val asset_to_return = shares * total_assets / total_supply

    if (asset_to_return == 0) {
      (Err("ZeroAmountNotAllowed"), state)
    } else {
      val user_balance = getOrElse(state.balances, info.sender, { amount: 0 }).amount

      if (user_balance < shares) {
        (Err("unsigned int error"), state)
      } else {
        val updated_user_balance = user_balance - shares
        val updated_total_supply = total_supply - shares

        val updated_balances = state.balances.put(info.sender, { amount: updated_user_balance })
        val updated_state = { ...state, config: { total_supply: updated_total_supply }, balances: updated_balances }

        val msg = CosmosMsg_Bank(BankMsg_Send({ to_address: info.sender, amount: [{ denom: DENOM, amount: asset_to_return }] }))

        (Ok(Response_new
          .add_attribute("action", FromStr("burn"))
          .add_attribute("user", FromStr(info.sender))
          .add_attribute("asset", FromInt(asset_to_return))
          .add_attribute("shares", FromInt(shares))
          .add_message(msg)), updated_state)
      }
    }
  }
```

### Changes Made

- Used `getOrElse` to safely retrieve the user's balance, providing a default balance of `{ amount: 0 }` if the user does not have an entry in the `state.balances` map. This prevents the `Called 'get' with a non-existing key` error.
\end{lstlisting}

The overall runtime repair process resembles semantic repair, because it relies on iterating over the user-provided I/O examples for detection.
However, the content that we use to prompt the LLM---primarily, error messages obtained by running Quint---is more akin to static repair rounds.
The only difference is that, when repairing static errors, we obtain error messages from the type checker as explained in Section~\ref{sec:static-errors}, and not by attempting to run the simulator (REPL) as we do when obtaining runtime error messages.
However, because of the aforementioned similarity in content, we prepare runtime repair prompts by reusing the static repair prompt template (shown in full in Listing~\ref{lst:full-repair-user-template-static}).

\section{Comprehensive model generation results}
\label{sec:additional-results}

\jan{In Table~\ref{tab:ctf-difficulty-3-study-3}, we present comprehensive results of our evaluation, including the complete outcomes for the Capture the Flag benchmark (CTF-02, 04, 05, 07, and 09).
    These results include information about both the generation process (\textbf{Avg. Repair}), and the testing that followed if the generated functions contained no runtime errors and were able to execute the tests (\textbf{Pass Rate}).}

\begin{table}[h]
    \centering
    \caption{Average number of repair rounds used per function and test pass rates for unseen test cases, averaged across $5$ independent runs (results for the whole CTF benchmark) \mynote{(Difficulty Level 3, Study 3)}}
    
    \begin{tabular}{
        |>{\centering\arraybackslash}p{1.9cm}
        >{\centering\arraybackslash}p{3cm}
        >{\centering\arraybackslash}p{3cm}
        >{\centering\arraybackslash}p{3cm}|}
    \hline
    \multicolumn{4}{|c|}{\underline{\textbf{CTF-02}}} \\
    \multicolumn{4}{|c|}{} \\[-0.8em]
    & \textbf{Function} & \textbf{Avg. Repair} & \textbf{Pass Rate} \\ \hline
    \cmark & \texttt{deposit} & 0.0 & 100\% \\
    \cmark & \texttt{withdraw} & 0.0 & 100\% \\
    \cmark & \texttt{stake} & 0.0 & 100\% \\
    \cmark & \texttt{unstake} & 0.0 & 50\% \\ \hline
    \end{tabular}

    \begin{tabular}{
        |>{\centering\arraybackslash}p{1.9cm}
        >{\centering\arraybackslash}p{3cm}
        >{\centering\arraybackslash}p{3cm}
        >{\centering\arraybackslash}p{3cm}|}
    \hline
    \multicolumn{4}{|c|}{\underline{\textbf{CTF-04}}} \\
    \multicolumn{4}{|c|}{} \\[-0.8em]
    & \textbf{Function} & \textbf{Avg. Repair} & \textbf{Pass Rate} \\ \hline
    \cmark & \texttt{instantiate} & 0.0 & 100\% \\
    \cmark & \texttt{mint} & 1.2 & 90\% \\
    \cmark & \texttt{burn} & 0.0 & 100\% \\ \hline
    \end{tabular}

    \begin{tabular}{
        |>{\centering\arraybackslash}p{1.9cm}
        >{\centering\arraybackslash}p{3cm}
        >{\centering\arraybackslash}p{3cm}
        >{\centering\arraybackslash}p{3cm}|}
    \hline
    \multicolumn{4}{|c|}{\underline{\textbf{CTF-05}}} \\
    \multicolumn{4}{|c|}{} \\[-0.8em]
    & \textbf{Function} & \textbf{Avg. Repair} & \textbf{Pass Rate} \\ \hline
    \cmark & \texttt{instantiate} & 0.0 & 100\% \\
    \cmark & \texttt{deposit} & 0.0 & 100\% \\
    \cmark & \texttt{withdraw} & 0.0 & 100\% \\
    \cmark & \texttt{owner\_action} & 0.0 & 100\% \\
    \cmark & \texttt{propose\_owner} & 0.0 & 100\% \\
    \cmark & \texttt{accept\_owner} & 0.0 & 100\% \\
    \cmark & \texttt{drop\_owner} & 0.0 & 100\% \\ \hline
    \end{tabular}

    \begin{tabular}{
        |>{\centering\arraybackslash}p{1.9cm}
        >{\centering\arraybackslash}p{3cm}
        >{\centering\arraybackslash}p{3cm}
        >{\centering\arraybackslash}p{3cm}|}
    \hline
    \multicolumn{4}{|c|}{\underline{\textbf{CTF-07}}} \\
    \multicolumn{4}{|c|}{} \\[-0.8em]
    & \textbf{Function} & \textbf{Avg. Repair} & \textbf{Pass Rate} \\ \hline
    \cmark & \texttt{instantiate} & 0.0 & 100\% \\
    \cmark & \texttt{deposit} & 0.8 & 100\% \\
    \cmark & \texttt{withdraw} & 0.0 & 100\% \\
    \cmark & \texttt{owner\_action} & 0.0 & 100\% \\
    \cmark & \texttt{update\_config} & 0.0 & 100\% \\ \hline
    \end{tabular}

    \begin{tabular}{
        |>{\centering\arraybackslash}p{1.9cm}
        >{\centering\arraybackslash}p{3cm}
        >{\centering\arraybackslash}p{3cm}
        >{\centering\arraybackslash}p{3cm}|}
    \hline
    \multicolumn{4}{|c|}{\underline{\textbf{CTF-09}}} \\
    \multicolumn{4}{|c|}{} \\[-0.8em]
    & \textbf{Function} & \textbf{Avg. Repair} & \textbf{Pass Rate} \\ \hline
    \cmark & \texttt{deposit} & 2.4 & 73\% \\
    \mixed \; (4/5) & \texttt{withdraw} & 1.6 & 73\% \\
    \cmark & \texttt{increase\_reward} & 0.2 & 100\% \\
    \mixed \; (4/5) & \texttt{claim\_rewards} & 1.8 & 80\% \\ \hline
    \end{tabular}

    \label{tab:ctf-difficulty-3-study-3}
\end{table}

\jan{Interestingly, the results for CTF-09 are the least promising.
    For two of its functions (\texttt{withdraw} and \texttt{claim\_rewards}), the generation process was successful in only $4$ out of $5$ runs.
    Moreover, this contract required the most rounds of repair on average, and has the worst pass rates on unseen test cases.
    The reason for this likely lies in the complex logic required for contract operations in CTF-09 (and, consequently, lengthy target Quint implementations).}

\jan{How do user-supplied natural language (NL) inputs, i.e., the English language description of each Quint action, affect performance across our CTF Benchmark?
    In order to answer this question, we also conducted an ablation study where we left out the NL descriptions when asking the LLM to generate or repair a function.
    We report the results of our ablation study in Table~\ref{tab:ctf-difficulty-5-study-3}.}

\begin{table}[h]
    \centering
    \caption{Ablation study with natural language descriptions left out \mynote{(Difficulty Level 5, Study 3)}}
    
    \begin{tabular}{
        |>{\centering\arraybackslash}p{1.9cm}
        >{\centering\arraybackslash}p{3cm}
        >{\centering\arraybackslash}p{3cm}
        >{\centering\arraybackslash}p{3cm}|}
    \hline
    \multicolumn{4}{|c|}{\underline{\textbf{CTF-02}}} \\
    \multicolumn{4}{|c|}{} \\[-0.8em]
    & \textbf{Function} & \textbf{Avg. Repair} & \textbf{Pass Rate} \\ \hline
    \cmark & \texttt{deposit} & 0.0 & 100\% \\
    \cmark & \texttt{withdraw} & 0.0 & 100\% \\
    \cmark & \texttt{stake} & 0.0 & 100\% \\
    \cmark & \texttt{unstake} & 0.0 & 50\% \\ \hline
    \end{tabular}

    \begin{tabular}{
        |>{\centering\arraybackslash}p{1.9cm}
        >{\centering\arraybackslash}p{3cm}
        >{\centering\arraybackslash}p{3cm}
        >{\centering\arraybackslash}p{3cm}|}
    \hline
    \multicolumn{4}{|c|}{\underline{\textbf{CTF-04}}} \\
    \multicolumn{4}{|c|}{} \\[-0.8em]
    & \textbf{Function} & \textbf{Avg. Repair} & \textbf{Pass Rate} \\ \hline
    \cmark & \texttt{instantiate} & 0.0 & 100\% \\
    \cmark & \texttt{mint} & 2.0 & 100\% \\
    \cmark & \texttt{burn} & 0.0 & 100\% \\ \hline
    \end{tabular}

    \begin{tabular}{
        |>{\centering\arraybackslash}p{1.9cm}
        >{\centering\arraybackslash}p{3cm}
        >{\centering\arraybackslash}p{3cm}
        >{\centering\arraybackslash}p{3cm}|}
    \hline
    \multicolumn{4}{|c|}{\underline{\textbf{CTF-05}}} \\
    \multicolumn{4}{|c|}{} \\[-0.8em]
    & \textbf{Function} & \textbf{Avg. Repair} & \textbf{Pass Rate} \\ \hline
    \cmark & \texttt{instantiate} & 0.0 & 100\% \\
    \cmark & \texttt{deposit} & 0.0 & 100\% \\
    \cmark & \texttt{withdraw} & 0.0 & 100\% \\
    \cmark & \texttt{owner\_action} & 0.0 & 100\% \\
    \cmark & \texttt{propose\_owner} & 0.0 & 100\% \\
    \cmark & \texttt{accept\_owner} & 0.0 & 100\% \\
    \cmark & \texttt{drop\_owner} & 0.0 & 100\% \\ \hline
    \end{tabular}

    \begin{tabular}{
        |>{\centering\arraybackslash}p{1.9cm}
        >{\centering\arraybackslash}p{3cm}
        >{\centering\arraybackslash}p{3cm}
        >{\centering\arraybackslash}p{3cm}|}
    \hline
    \multicolumn{4}{|c|}{\underline{\textbf{CTF-07}}} \\
    \multicolumn{4}{|c|}{} \\[-0.8em]
    & \textbf{Function} & \textbf{Avg. Repair} & \textbf{Pass Rate} \\ \hline
    \cmark & \texttt{instantiate} & 0.0 & 100\% \\
    \cmark & \texttt{deposit} & 1.0 & 100\% \\
    \cmark & \texttt{withdraw} & 0.2 & 100\% \\
    \cmark & \texttt{owner\_action} & 0.0 & 100\% \\
    \cmark & \texttt{update\_config} & 0.0 & 100\% \\ \hline
    \end{tabular}

    \begin{tabular}{
        |>{\centering\arraybackslash}p{1.9cm}
        >{\centering\arraybackslash}p{3cm}
        >{\centering\arraybackslash}p{3cm}
        >{\centering\arraybackslash}p{3cm}|}
    \hline
    \multicolumn{4}{|c|}{\underline{\textbf{CTF-09}}} \\
    \multicolumn{4}{|c|}{} \\[-0.8em]
    & \textbf{Function} & \textbf{Avg. Repair} & \textbf{Pass Rate} \\ \hline
    \mixed \; (4/5) & \texttt{deposit} & 2.0 & 67\% \\
    \mixed \; (2/5) & \texttt{withdraw} & 2.4 & 40\% \\
    \cmark & \texttt{increase\_reward} & 0.8 & 100\% \\
    \cmark & \texttt{claim\_rewards} & 1.8 & 100\% \\ \hline
    \end{tabular}

    \label{tab:ctf-difficulty-5-study-3}
\end{table}

\jan{As can be seen, our results show a low impact of NL user inputs on the performance across our benchmark, suggesting that the necessary level of user expertise and involvement is lower than originally anticipated for a task of this level of complexity.
    When user NL inputs are withheld, CTF-04, 07, and 09 required a slightly higher average number of repair rounds, while performance on CTF-02 and 05 was unaffected.}

\jan{Notably, performance was more negatively affected on CTF-09, which includes the most challenging Quint functions.
    Although firm conclusions are limited by the scope of this ablation study, these results suggest that custom NL user inputs may play a more significant role in more demanding generation tasks.}

Finally, in Figure~\ref{fig:listings-stubs}, we provide an illustrative example showcasing the progression from the original Rust code to the corresponding Quint stub and the final LLM-generated Quint code, demonstrating the translation and generation process for a function from the CTF-02 benchmark.

\begin{figure}[h]
\begin{lstlisting}[language=Rust, style=ruststyle, caption={Original Rust code}]
/// Entry point for user to stake tokens
pub fn deposit(deps: DepsMut, info: MessageInfo) -> Result<Response, ContractError> {
    // validate denom
    let amount = must_pay(&info, DENOM).unwrap();

    // increase total stake
    let mut user = VOTING_POWER
        .load(deps.storage, &info.sender)
        .unwrap_or_default();
    user.total_tokens += amount;

    VOTING_POWER
        .save(deps.storage, &info.sender, &user)
        .unwrap();

    Ok(Response::new()
        .add_attribute("action", "deposit")
        .add_attribute("user", info.sender)
        .add_attribute("amount", amount))
}
\end{lstlisting}

\begin{lstlisting}[language=quint, style=quintstyle, caption={Quint stub}]
pure def deposit(state: ContractState, deps: Deps, info: MessageInfo): (Result[Response, ContractError], ContractState) = {
    // TODO: Update body
    (Ok(Response_new), state)
}
\end{lstlisting}

\begin{lstlisting}[language=quint, style=quintstyle, caption={LLM-generated Quint code}]
pure def deposit(state: ContractState, info: MessageInfo): (Result[Response, ContractError], ContractState) = {
    match must_pay(info, DENOM) {
        | Err(e) => (Err(e), state)
        | Ok(amount) => {
        val user = getOrElse(state.voting_power, info.sender, { total_tokens: 0, voting_power: 0, released_time: 0 })
        val updated_user = { ...user, total_tokens: user.total_tokens + amount }
        val updated_voting_power = state.voting_power.put(info.sender, updated_user)
        val new_state = { ...state, voting_power: updated_voting_power }
        
        (Ok(Response_new
            .add_attribute("action", FromStr("deposit"))
            .add_attribute("user", FromStr(info.sender))
            .add_attribute("amount", FromInt(amount))), new_state)
        }
    }
    }
\end{lstlisting}

\caption{Example of a stub along with the original Rust code that induced it and the final LLM-generated Quint code. Source: CTF-02.}
\label{fig:listings-stubs}
\end{figure}

\section{Adapter Generation and Repair}
\label{sec:adapter-details}

While mechanically generating a Quint model for a smart contract, we also generate an adapter stub, a Rust test file which can run traces induces by the Quint model against the CosmWasm implementation of the smart contract.
Afterwards, we use an LLM to improve the adapter by regenerating a core part of the test's functionality: the state comparison function \texttt{compare\_state}.
In Listing~\ref{lst:compare-state-stub}, we present the stub for the \texttt{compare\_state} function (the mechanically generated default with weak checks), and in Listings \ref{lst:compare-state-result-ctf-02} and \ref{lst:compare-state-result-ctf-04} we present example generations from our study in Appendix~\ref{sec:case-study-mbt} for CTF-02 and 04, respectively.


\begin{lstlisting}[language=Rust, label=lst:compare-state-stub, style=ruststyle, caption={Stub used for the state comparison functions during adapter generation}]
fn compare_state(test_state: &TestState, app: &App, state: &State) {
        // compare contract balances
        let balance = app
            .wrap()
            .query_balance(&test_state.contract_addr, DENOM)
            .unwrap()
            .amount;
        let trace_balance = state
            .bank
            .get(&test_state.contract_addr.to_string())
            .and_then(|x| x.get(DENOM))
            .and_then(|x| x.to_u128())
            .unwrap_or(0);
        println!(
            "Contract balance ({:?}) for {DENOM}: {:?} vs {:?}",
            test_state.contract_addr,
            balance,
            Uint128::new(trace_balance)
        );
        assert_eq!(balance, Uint128::new(trace_balance));

        // TODO: Query the contract and compare the state as you wish
    }
\end{lstlisting}



\begin{lstlisting}[language=Rust, label=lst:compare-state-result-ctf-02, style=ruststyle, caption={CTF-02 generated state comparison example}]
fn compare_state(test_state: &TestState, app: &App, state: &State) {
    // compare contract balances
    let balance = app
        .wrap()
        .query_balance(&test_state.contract_addr, DENOM)
        .unwrap()
        .amount;
    let trace_balance = state
        .bank
        .get(&test_state.contract_addr.to_string())
        .and_then(|x| x.get(DENOM))
        .and_then(|x| x.to_u128())
        .unwrap_or(0);
    println!(
        "Contract balance ({:?}) for {DENOM}: {:?} vs {:?}",
        test_state.contract_addr,
        balance,
        Uint128::new(trace_balance)
    );
    assert_eq!(balance, Uint128::new(trace_balance));

    // compare contract state
    for (user, trace_user_info) in &state.contract_state.voting_power {
        // Query user info
        let msg = oaksecurity_cosmwasm_ctf_02::msg::QueryMsg::GetUser {
            user: user.clone(),
        };
        let user_info: oaksecurity_cosmwasm_ctf_02::state::UserInfo = app
            .wrap()
            .query_wasm_smart(test_state.contract_addr.to_owned(), &msg)
            .unwrap();

        // Compare total tokens
        let trace_total_tokens = Uint128::new(trace_user_info.total_tokens.to_u128().unwrap());
        assert_eq!(user_info.total_tokens, trace_total_tokens, "Mismatch in total tokens for user {}", user);

        // Compare voting power
        let trace_voting_power = trace_user_info.voting_power.to_u128().unwrap();
        assert_eq!(user_info.voting_power, trace_voting_power, "Mismatch in voting power for user {}", user);

        // Compare release time
        let trace_released_time = trace_user_info.released_time.to_u64().unwrap();
        assert_eq!(user_info.released_time.seconds(), trace_released_time, "Mismatch in released time for user {}", user);

        // Query voting power separately
        let msg = oaksecurity_cosmwasm_ctf_02::msg::QueryMsg::GetVotingPower {
            user: user.clone(),
        };
        let voting_power: u128 = app
            .wrap()
            .query_wasm_smart(test_state.contract_addr.to_owned(), &msg)
            .unwrap();

        assert_eq!(voting_power, trace_voting_power, "Mismatch in voting power query for user {}", user);
    }
}
\end{lstlisting}

\begin{lstlisting}[language=Rust, label=lst:compare-state-result-ctf-04, style=ruststyle, caption={CTF-04 generated state comparison example}]
fn compare_state(test_state: &TestState, app: &App, state: &State) {
    // compare contract balances
    let balance = app
        .wrap()
        .query_balance(&test_state.contract_addr, DENOM)
        .unwrap()
        .amount;
    let trace_balance = state
        .bank
        .get(&test_state.contract_addr.to_string())
        .and_then(|x| x.get(DENOM))
        .and_then(|x| x.to_u128())
        .unwrap_or(0);
    println!(
        "Contract balance ({:?}) for {DENOM}: {:?} vs {:?}",
        test_state.contract_addr,
        balance,
        Uint128::new(trace_balance)
    );
    assert_eq!(balance, Uint128::new(trace_balance));

    // Query the contract's config
    let config_msg = oaksecurity_cosmwasm_ctf_04::msg::QueryMsg::GetConfig {};
    let contract_config: oaksecurity_cosmwasm_ctf_04::state::Config = app
        .wrap()
        .query_wasm_smart(test_state.contract_addr.to_owned(), &config_msg)
        .unwrap();
    let trace_config = &state.contract_state.config;
    assert_eq!(
        contract_config.total_supply,
        Uint128::new(trace_config.total_supply.to_u128().unwrap())
    );

    // Query the contract's user balances
    for (address, trace_balance) in &state.contract_state.balances {
        let balance_msg = oaksecurity_cosmwasm_ctf_04::msg::QueryMsg::UserBalance {
            address: address.clone(),
        };
        let contract_balance: oaksecurity_cosmwasm_ctf_04::state::Balance = app
            .wrap()
            .query_wasm_smart(test_state.contract_addr.to_owned(), &balance_msg)
            .unwrap();
        assert_eq!(
            contract_balance.amount,
            Uint128::new(trace_balance.amount.to_u128().unwrap())
        );
    }
}
\end{lstlisting}


\subsection{Adapter generation testing and repair results}

As part of our model-based testing case study from Appendix~\ref{sec:case-study-mbt}, 
we evaluated of our adapter generation method on CTF-02 and 04.
For each of the two adapters, we only use the LLM to generate one particular (Rust) function: \texttt{compare\_state}.
After (re-)generating this function, we attempt to run the whole test using the \text{cargo test} utility.
Given that the LLM may also produce errors when generating Rust code, the resulting adapter can fail to compile and execute.
Therefore, implementing a repair loop is necessary here as well.
In Table~\ref{tab:adapter-results} we summarize the full results of the generate-and-repair loop, obtained throughout our model-based testing case study.

\begin{table}[h]
    \centering
    \begin{tabular}{|c|c|c|c|}
        \hline
        \textbf{CTF} & \textbf{Adapter Function} & \textbf{Avg. Repair} & \textbf{Static Rounds} \\
        \hline
        CTF-02 & \texttt{compare\_state} & 0.40 & 2 \\
        CTF-04 & \texttt{compare\_state} & 0.00 & 0 \\
        \hline
    \end{tabular}
    \caption{Results for adapter generation for CTF-02 and CTF-04 across $5$ independent runs}
    \label{tab:adapter-results}
\end{table}

To summarize, our adapter generation method resulted in a working adapter test in all $5$ runs, and we only required two rounds of repair in total.
However, it must be noted that, in this case, mechanical generation carries an even larger portion of the responsibility: the LLM's task is to simply expand the checks carried out by a single Rust function (a language which LLMs handle with higher proficiency than Quint).

\section{Prompt Templates}
\label{sec:prompt-templates}

In this section, we present an informative selection of prompt templates that we use to format the system's LLM queries, and to create our few-shot demonstrations dataset.
The full list of templates is available in the project's repository on GitHub.\footnote{\url{https://github.com/informalsystems/FuzzMo/tree/main/data/prompts}}

\subsection{Full model generation user prompt template}
\label{sec:full-user-template}

\begin{lstlisting}[language=, style=templatestyle,label=lst:full-user-template, caption={User prompt template}]
Please complete this stub for my `@@@NAME@@@` Quint function.

@@@DESCRIPTION@@@

```
@@@STUB@@@
```

Here is the contract state type and related types:

```
@@@QUINT TYPE DEFINITIONS@@@
```

The following constants are available to the function:

```
@@@CONSTANTS@@@
```
@@@DEC@@@

The following Quint imports are available:

```
@@@QUINT IMPORTS@@@
```

If a module is imported as `import Module.* from "./lib/module"`, you must use the identifiers from it directly without (so `name` instead of `Module.name`).

The `@@@NAME@@@` Quint function models a CosmWasm Rust function for processing smart contract messages.
@@@MESSAGE HANDLERS@@@

Your Quint code must be functional.
All function arguments are immutable, and you can not use mutable variables.
@@@IO EXAMPLES@@@
\end{lstlisting}

\subsection{Full repair user prompt template}
\label{sec:full-repair-user-template}







Listing~\ref{lst:full-repair-user-template-static} presents the full prompt template we utilize for generating user messages during the static error repair phase.
The macro \texttt{\textcolor{templatekeyword}{@@@ORIGINAL IMPLEMENTATION@@@}} expands to a stub implementation that is being repaired (i.e., the output of the initial code generation or a preceding repair round), and \texttt{\textcolor{templatekeyword}{@@@QUINT ERRORS@@@}} expands to the output of the Quint type checker (or, in the case of a runtime repair round, the standard output of the Quint executable).

\begin{lstlisting}[language=, style=templatestyle,label=lst:full-repair-user-template-static, caption={Static and runtime repair user prompt template}]
Please repair my `@@@NAME@@@` Quint function so that it no longer has Quint errors.

@@@DESCRIPTION@@@

Here is the contract state type and related types:

```
@@@QUINT TYPE DEFINITIONS@@@
```

The following constants are available to the function:

```
@@@CONSTANTS@@@
```
@@@DEC@@@

@@@QUINT SHORT INSTRUCTIONS@@@

Here is the current implementation:

```
@@@ORIGINAL IMPLEMENTATION@@@
```

Here are the Quint errors that the current implementation has:

@@@QUINT ERRORS@@@

Ignore problems that do not concern this function.
Only respond with a function definition (`pure def`), and not type definitions.

First explain what the problem is with a bullet-point list.
Do not make guesses about other problems.
Then respond with the repaired function that does not have those errors.
Minimize your changes to the original implementation, and please respect Quint rules.
Do not introduce your own type definitions.
\end{lstlisting}

\section{Quint Model Structure}
\label{sec:quint-model-structure}

In this section, we give a basic overview of how a Quint model looks like.
This is far from a complete description, but it should give you enough detail to understand the decisions made when generating Quint stubs.
For a more detailed description, refer to the Quint documentation~\cite{quint}.

Quint is a modelling language. 
It is like a very familiar pseudocode, that additionally enables us to encode state machines (the evolution of the system) and nondeterministic choices.

The state of the Quint model is defined by a set of variables, for example the following:

\begin{lstlisting}[language=quint]
    var x: int
    var y: bool
\end{lstlisting}

The model's evolution is described by its initial state and possible transitions from each state.
In Quint, there is a special kind of function, called \lstinline|action|, that can modify state variables.
Thus, we use actions to define both the initial state and transitions.

As an example, the initial state may be defined as follows:
\begin{lstlisting}[language=quint]
    action init = all {
        x' = 0,
        y' = false
    }
\end{lstlisting}

And transitions may be defined like this:
\begin{lstlisting}[language=quint]
    action step = any {
        nondet increase = oneOf(1.to(10))
        all{
            x < 5,
            y == false,
            x' = x + increase,
            y' = true
        },
        all{            
            y' = true,
            x' = 0
        }        
    }
\end{lstlisting}

Let us explain the above two actions.
The syntax \lstinline[language=quint]|x'| (read: \emph{x primed}) means ``the value of \lstinline[language=quint]|x| in the next state''.
Thus, the \lstinline[language=quint]|action step| defines two possible transitions:
\begin{itemize}
    \item \textbf{either} \lstinline[language=quint]|x < 5| and \lstinline[language=quint]|y == false|, 
    in which case increase \lstinline[language=quint]|x| in the next state by a non-deterministically chosen number between 1 and 10,
    and set \lstinline[language=quint]|y| to \lstinline|true|,
    \item \textbf{or} set \lstinline[language=quint]|y| to \lstinline|true| and \lstinline[language=quint]|x| to 0.
\end{itemize}
The \emph{either-or} choice is encoded by the \lstinline[language=quint]|any| keyword.
Furthermore, which one of them is chosen is non-deterministic, 
but the first option can only be selected if its conditions are satisfied.

The \lstinline[language=quint]|action init| is straightforward: 
it sets the initial values of \lstinline[language=quint]|x| and \lstinline[language=quint]|y| to 0 and \lstinline[language=quint]|false|, respectively.

Beside actions, Quint has a concept of \emph{pure functions}.
Those are the functions that may not modify the state, but can be used to calculate values.

The last point to mention is the concept of invariants.
They are defined on state variables, and Quint's simulator or model checker make sure that they hold at all states of the model evolution.
As an example of an invariant, consider the following simple one:
\begin{lstlisting}[language=quint]
    val xAlwaysSmall = invariant x < 15
\end{lstlisting}

\end{document}